# Nonlinear dielectric geometric-phase metasurface with simultaneous structure and lattice symmetry design

Bingyi Liu[1,*], René Geromel[2,*], Zhaoxian Su[3], Kai Guo[1], Yongtian Wang[3], Zhongyi Guo[1], Lingling Huang[3,†], and Thomas Zentgraf[2,4‡]

[1]*School of Computer Science and Information Engineering, Hefei University of Technology, Hefei, 230009, China*
[2]*Department of Physics, Paderborn University, Paderborn, 33098, Germany*
[3]*School of Optics and Photonics, Beijing Engineering Research Center of Mixed Reality and Advanced Display, Beijing Institute of Technology, Beijing, 100081, China*
[4]*Institute for Photonic Quantum Systems, Paderborn University, Paderborn, 33098, Germany*

**Abstract**

In this work, we utilize thin dielectric meta-atoms placed on a silver substrate to efficiently enhance and manipulate the third harmonic generation. We theoretically and experimentally reveal that when the structural symmetry of the meta-atom is incompatible with the lattice symmetry of an array, some generalized nonlinear geometric phases appear, which offers new possibilities for harmonic generation control beyond the accessible symmetries governed by the selection rule. The underlying mechanism is attributed to the modified rotation of the effective principal axis of a dense meta-atom array, where the strong coupling among the units gives rise to a generalized linear geometric phase modulation on the pump light. Therefore, nonlinear geometric phases carried by the third-harmonic emissions are the natural result of the wave-mixing process among the modes excited at the fundamental frequency. This mechanism further points out a new strategy to predict the nonlinear geometric phases delivered by the nanostructures according to their linear responses. Our design is simple and efficient, and offers alternatives for the nonlinear meta-devices that are capable of flexible photon generation and manipulation.

---

[*] These two authors contributed equally to this work.
[†] Email: huanglingling@bit.edu.cn
[‡] Email: thomas.zentgraf@uni-paderborn.de

Metasurfaces, those that are made of elaborately designed artificial nanostructures, pave the way to comprehensive light propagation control with an ultra-compact platform.[1,2] Benefiting from the collective oscillation of electrons in the vicinity of the surface of plasmonic structures[3,4] or the multipolar Mie resonances supported by high-index nanoparticles,[5,6] light waves can be confined down to the nanoscale for strongly enhanced light-matter interactions. Due to the intense scattering of the field trapped under the resonance conditions, exquisite manipulation of the key features of light is possible by tailoring the dispersion properties of artificial meta-atoms, which further endows the metasurfaces with numerous functionalities, such as imaging,[7,8] holographic display,[9,10] image edge detection,[11] high-capacity optical information storage and encryptions,[12,13] to name a few.

Nonlinear optical phenomena in metasurfaces, e.g., harmonic generations, are also boosted due to the less rigorous requirement of the phase-matching condition.[14-24] Also, the manipulation of the nonlinear phase attracted tremendous attention in recent years since it allows for easy control of nonlinear wavefronts. Like its linear counterpart, both the nonlinear resonant phase[25-28] and the nonlinear Pancharatnam-Berry phase, namely, the nonlinear geometric phase,[29-36] are two major solutions to spatial nonlinearity engineering. It is worth mentioning that nonlinear geometric-phase metasurfaces facilitate the functional nonlinear beam shaping by locally varying the orientation angles of the nanoantennas, by which the complex spin-rotation coupling effect gives rise to the spin-dependent nonlinear geometric-phase terms.

To date, there are mainly two strategies for the realization of simultaneous efficient harmonic generations and nonlinear geometric-phase manipulation in metasurfaces: plasmonic-nonlinear active medium hybrid structures, such as multi-quantum-wells,[32,33,37] PFO coating,[30,31] 2D materials,[38] thin ITO layers,[39] and all-dielectric structures, such as silicon.[35,36,40,41] The reported works that involve the plasmonic structures show apparent drawbacks because plasmonic materials cannot be used under high-power operation due to their low laser-induced damage threshold, and their nonlinear conversion efficiency is also limited by high intrinsic Ohmic loss and the small mode volume. On the contrary, high index all-dielectric structures possess the merit of low loss, larger mode volume and higher damage threshold, and thereby potentially facilitate efficient frequency conversions.[42-51]

Recently, replacing the transparent insulator substrate with a near-zero-index substrate[20] or a highly reflective substrate,[52] has been revealed to effectively improve the quality factors of the micro-resonators, where extremely strong resonant modes can be excited and giant local field enhancements can be obtained. Inspired by this property, here, we utilize thin dielectric meta-atoms placed on a metallic substrate (i.e., Ag) to efficiently reflect and improve the intensity of third-harmonic (TH) emissions. Moreover, flexible nonlinear geometric phases are available by tuning both the symmetry and the local orientation angle of the meta-atoms. Our design shows at least three orders of magnitude improvement of the nonlinear conversion efficiency when compared with the same structures placed on a transparent substrate. In addition, the small aspect ratio associated with the thin profile, whose typical thickness is about 1/12 of the fundamental wavelength (~1200 nm), greatly facilitates sample fabrication. In addition to the meta-atoms of one-fold (C1), two-fold (C2) or four-fold (C4) rotational symmetries, here we further theoretically and experimentally reveal that the meta-atoms of three-fold (C3) symmetry could also provide a generalized geometric-phase modulation when they are placed in a square lattice, and C4 meta-atom placed in hexagonal lattice could also maintain an apparent nonlinear geometric-phase modulation on the co-polarized TH conversion process. The underlying mechanism is understood as the strong optical anisotropy appearing at the fundamental frequency, i.e., a dense meta-atom array with a given local orientation angle would introduce a modified rotation of its effective optical principal axis, which is the origin of generalized linear geometric phase modulation on the pump light. Then, based on the wave-mixing process involving the modes excited at the fundamental frequency, TH signals that carry nontrivial phase terms are observed, i.e., the generalized nonlinear geometric phases. Therefore, specific combinations of the structure symmetries and lattice symmetries can offer new possibilities in nonlinear optical process control.

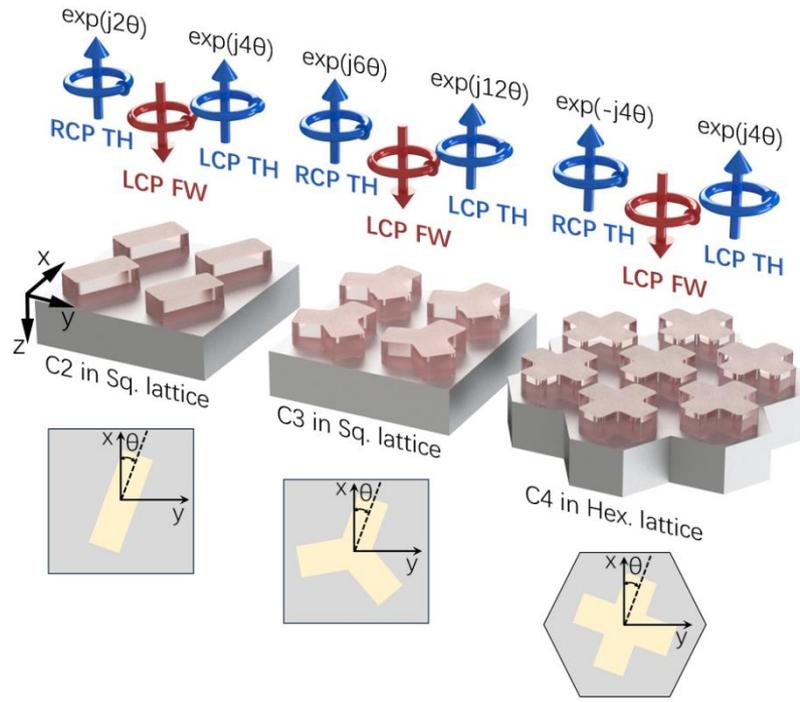

**Figure 1.** Third harmonic generation in a thin dielectric meta-atom array with tailored structure and lattice symmetry.

## RESULTS AND DISCUSSION

### *Wave-mixing picture for the interpretation of the generalized nonlinear geometric phase beyond the selection rule*

Figure 1 shows the schematic of the TH emissions from C2/C3 meta-atoms in a square lattice and C4 meta-atoms in a hexagonal lattice with a given local orientation angle $\theta$. As depicted in the figure, by varying $\theta$ from 0° to 360°, the TH signal emitted to the upper half-space possesses a phase term proportional to multiples of $\theta$, which is known as the nonlinear geometric phase. However, here we observe some new cases beyond the nonlinear optical selection rule of artificial nanostructure lattice.[31] For example, the C3 meta-atoms in a square lattice could support specific nonlinear geometric phase modulations on emitted TH signals, and the C4 meta-atoms in a hexagonal lattice could also deliver co-polarized TH emission to the far-field while showing obvious phase modulation. The appearance of the above nontrivial nonlinear geometric phases can be interpreted from the nonlinear polarizations induced inside the dielectric meta-atoms, where the complex mode coupling process should be considered rather than dipole approximation in thin metallic meta-atoms.[31,53] As reported in our previous work,[36] nonlinear polarizations in silicon meta-atoms can be analytically derived from the mode coupling, or equivalently, the wave-mixing process among different modes excited by the pump light illumination. The equivalent rotation of the fundamental modes in circular polarization representation (CPR) gives rise to a linear geometric phase, which is exactly the base of the following nonlinear geometric phase. In this work, the scenario is slightly different because the pump light and TH emissions locate at the same half-space. Therefore, we define the forward circular polarization representation (FCPR) and backward circular polarization representation (BCPR) to help describe the interactions between the pump light and meta-atoms. In addition, the linear polarization representation (LPR) is fixed as a reference, which is exactly a forward linear polarization representation (FLPR).

In this work, the nonlinear phase modulation is characterized by an effective third-order nonlinear susceptibility tensor of the meta-atom array before (without prime) and after (with prime) varying its local orientation angle, which is analytically given as:

$$\chi^{(3)}_{\alpha'\beta'\gamma'\delta'} = \sum_{\alpha,\beta,\gamma,\delta} {}^{-}R_\alpha{}^{\alpha'} \chi^{(3)}_{\alpha\beta\gamma\delta} \left[ {}^{+}R_\beta{}^{\beta'} \right]^{-1} \left[ {}^{+}R_\gamma{}^{\gamma'} \right]^{-1} \left[ {}^{+}R_\delta{}^{\delta'} \right]^{-1} \quad (1)$$

Here, $\chi^{(3)}_{\alpha\beta\gamma\delta}$ is the third-order nonlinear susceptibility, $\alpha$, $\beta$, $\gamma$, and $\delta$ refer to the base of CPR, i.e., $L$, $R$, and $z$, and they correspond to the value of 1, –1 and 0 in the rotation element ${}^{\pm}R_\alpha{}^{\alpha'} = \delta_{\alpha'\alpha} \exp(\mp j\alpha\theta_{eff})$, where ± corresponds to FCPR and BCPR, and $\delta_{\alpha'\alpha}$ is Kronecker delta. In particular, $\theta_{eff}$ describes the optical principal axis rotation of the effective medium formed by a periodic meta-atom array, and it is substantially determined by the coupling strength among the meta-atoms. Detailed derivation of Eq. (1) can be found in Supporting Information.

According to Eq. (1), the rotation elements are the base of the nonlinear phase acquired from the wave-mixing process, therefore, the key to determine the possible nonlinear geometric phases is simplified to figure out the rotation element. Generally, considering an isolated meta-atom or meta-atom array with simultaneous strong local resonance and weak inter-structure coupling strength, the rotation angle of the effective principal axis is exactly equal to that of the structure, i.e., $\theta_{eff} = \theta$. However, the strong coupling among the dense meta-atoms introduces significant optical anisotropy, which is characterized by a modified effective optical principal axis. In this scenario, the orientation angle of a meta-atom is usually not the same as the rotation angle of the effective optical principal axis, and the general relation between $\theta_{eff}$ and $\theta$ can be characterized by an integer $l$, i.e., $\theta_{eff} = l\theta$. Therefore, the rigorous phase modulation associated with the modes excited at the fundamental frequency should be a generalized linear geometric phase.[54,55] In Supporting Information, we discuss the appearance of the linear geometric phase via the picture of emissions from linear polarizations, which is consistent with the picture that harmonic emissions are enabled by the corresponding nonlinear polarizations. Based on this, we determine the rotation elements according to the linear response of the meta-atom array at the fundamental frequency, and then we could further predict the allowed nonlinear geometric phases via Eq. (1).

### *Nonlinear geometric phase obtained with C2/C4/C3 meta-atoms placed in square lattice*

We first theoretically investigate the third harmonic generation (THG) of silicon C2/C4 meta-atom placed in a square lattice, which conforms to the selection rule in artificial nanostructure lattice.[31] Next, we further demonstrate that silicon C3 meta-atoms placed in a square lattice can also deliver TH emissions into the far-field while showing good nonlinear geometric phase modulation. Although it is inconsistent with the selection rule, it can be well-explained with the wave-mixing model and the nonlinear geometric phases can be analytically derived from Eq. (1). In our theoretical design, the thickness of the silicon meta-atoms is optimized as 105 nm, and the metallic substrate, i.e., an Ag film with a thickness of 200 nm is optically thick, which can efficiently reflect the pump light like a perfect electric conductor. We utilize the commercial finite element method (FEM) solver COMSOL Multiphysics wave optics module to help investigate the THG in thin silicon meta-atoms. Based on the electric fields $\mathbf{E}(\mathbf{r},\omega)$ that are obtained by normally illuminating the periodic silicon meta-atom array with a circularly polarized fundamental wave, the TH response is calculated based on the induced nonlinear polarization $\mathbf{P}_{NL}^{(3)}(\mathbf{r},3\omega) = \varepsilon_0 \chi^{(3)} : \mathbf{EEE}$ or nonlinear currents $\mathbf{J}_{NL}^{(3)}(\mathbf{r},3\omega) = j\omega\mathbf{P}_{NL}^{(3)}$, here $\mathbf{r}$ and ω refer to the spatial position in the meta-atoms and the angular frequency of the fundamental wave, respectively. The third-order nonlinear susceptibility of amorphous silicon (α-Si) used in our simulation is $\chi_{\alpha-Si}^{(3)} = 2.45\times10^{-19}\,\mathrm{m^2/V^2}$,[27] and the complex refractive index of α-Si and Ag is retrieved from experiment data.[36,56] However, the TH emission contributed by the Ag film is weak and is neglected in this work. Notably, the surface plasmon in Ag film plays an important role in trapping the fundamental wave and enhancing the nonlinear conversion efficiency, see Supporting Information for more details, where we compare the TH emissions from the Ag film, Au film and perfect electric conductor substrate.

The left panel of Figure 2a shows the schematic of silicon meta-atom of C2 rotational symmetry. A left-handed circularly polarized fundamental wave, whose wavelength is 1250 nm, is normally illuminated on the meta-atom array, of which the electric field amplitude is $10^8\,\mathrm{V/m}$ and the corresponding intensity is about $1.33\,\mathrm{GW/cm^2}$. The length and width of the C2 silicon meta-atom are 300 nm and 200 nm, respectively. In addition, a periodic boundary condition is applied along the x and y direction. Further, we replace the sharp corners of the structure with a fillet geometry to avoid the intense field concentration appeared in numerical calculation. A uniform field distribution of left-handed circularly polarized (LCP) and right-handed circularly polarized (RCP) components of the TH signals is obtained with the constraint that the meta-atom period is smaller than the TH wavelength. To be consistent with the polarization conversion process defined in the transmissive scenario, here we denote the case of LCP fundamental wave input and LCP TH wave emission (LCP to LCP) as a cross-polarized conversion process, and vice versa.[32,57] The averaged conversion efficiencies of co-polarized and cross-polarized TH signals are $6.5\times10^{-8}$ and $2\times10^{-8}$, respectively, which are even higher than previous thick transmissive structures of optimized thickness.[36] Here, the nonlinear conversion efficiency is defined by the ratio between the power of TH emissions and incident fundamental wave. We obtain a significant improvement of the nonlinear conversion efficiency by 3 to 4 orders of magnitude when compared with the same structure placed on a silica substrate, whose averaged conversion efficiencies of co-polarized and cross-polarized TH signals are $1.1\times10^{-11}$ and $8\times10^{-13}$, respectively. By varying the local orientation angle of the periodic meta-atom array, the nonlinear geometric phases obtained with C2 silicon meta-atom placed in a square lattice are proportional to $\exp(j\sigma 2\theta)$ and $\exp(j\sigma 4\theta)$ for the co-polarized and cross-polarized conversion (Figure 2a). Here, $\sigma$ is the spin of the incident fundamental wave, and $\theta$ is the local orientation angle of meta-atom. In this work, we generally assume the fundamental wave is an LCP wave, i.e., $\sigma=1$.

For C4 silicon meta-atoms placed in a square lattice, its nonlinear response can be understood as the superposition of the nonlinear emissions of two sets of C2 units (length: 330 nm, width: 130 nm) whose included angle is 90°. Therefore, the far-field co-polarized TH wave would disappear due to the destructive interference. In this scenario, the co-polarized TH energy is trapped in the vicinity of the meta-atom array, see left panel of Figure 2b. On the contrary, the cross-polarized conversion process possesses the nonlinear geometric phase of $\exp(j4\theta)$, and its nonlinear conversion efficiency is around $2\times10^{-8}$ (Figure 2b), which shows apparent improvement when compared with an identical structure placed on a silicon dioxide substrate, i.e., $4.7\times10^{-13}$. Up to now, the nonlinear geometric phases obtained with C2 and C4 meta-atoms placed in a square lattice agree well with the nonlinear optical selection rule.

Next, we investigate the THG in a C3 silicon meta-atom placed in a square lattice, see left panel of Figure 2c. Here, the C3 structure can be decomposed into three C2 structures (length: 170 nm, width: 150 nm) with equal included angle of 120° around its midpoint. The selected fundamental wavelength, period and thickness of meta-atom are 1200 nm, 390 nm, and 105 nm, respectively. The far-field co-polarized TH signal shows the geometric phase of $\exp(j6\theta)$ (Figure 2c) and its averaged TH conversion efficiency is $3.5\times10^{-9}$, which is improved by 3 orders of magnitude when compared with the same C3 structure placed on a silica substrate, i.e., $6.0\times10^{-13}$. In addition, the far-field cross-polarized TH signal also shows periodic phase modulation but fails to cover the $2\pi$ range (Figure 2c).

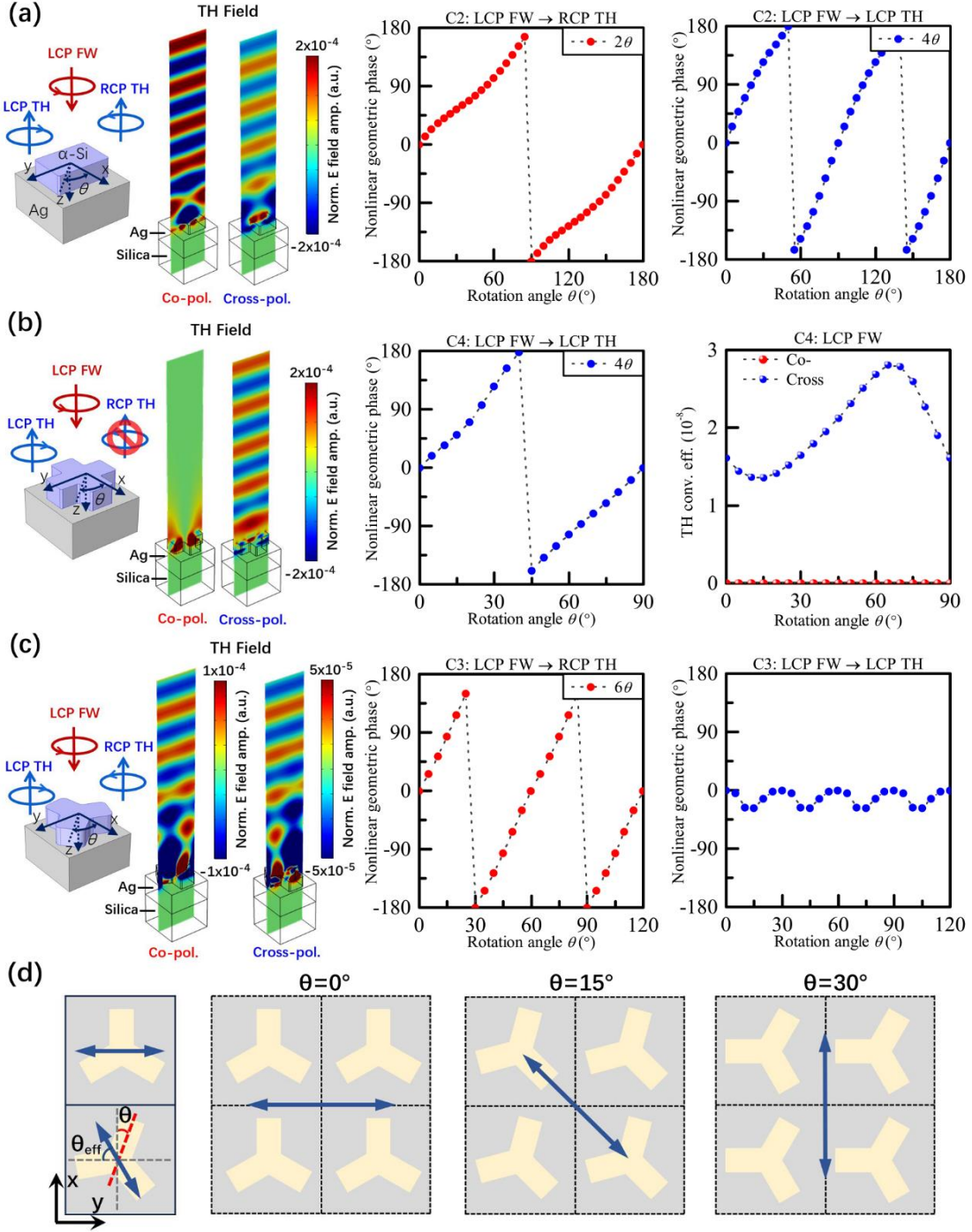

**Figure 2.** THG of thin dielectric meta-atom placed in a square lattice. (a) TH fields and nonlinear geometric phase modulation of C2 meta-atom. (b) TH fields, nonlinear geometric phase, and TH conversion efficiency of C4 meta-atom. (c) TH fields and nonlinear geometric phase of C3 meta-atom. (d) Rotation of effective principal optical axis (blue arrow) when varying the local orientation angle of C3 meta-atom array (characterized by red dashed line).

In order to explain the nonlinear phase modulation obtained with C3 meta-atom array, we need to revisit the selection rule in nonlinear metasurfaces. The nonlinear processes in metasurfaces, which are prohibited by the selection rule, indeed refer to the case that the nonlinear optical signals generated in the materials cannot propagate to the far-field. In other words, the material is excited by the pump light to produce the TH emissions but most of the TH energy is trapped in the vicinity of the meta-atom array and dissipates as the Joule heat. Notably, this nature offers us another possibility that if we can couple out the trapped TH energy to the far-field, then we have the chance to break the selection rule of nonlinear metasurfaces.

Previously reported work on the selection rule of nonlinear optical processes in C3 plasmonic meta-atom generally focuses on the manipulation of second harmonic generation (SHG) while its TH emission is prohibited.[29,30] However, for C3 silicon meta-atom placed in square lattice, the TH energy trapped in the lattice can be coupled out to the free-space via the scattering caused by the symmetry mismatch between structure and lattice. Based on this, the nonlinear geometric phases associated with co-polarized and cross-polarized nonlinear conversion are understood via a wave-mixing picture. We first determine the rotation angle of the effective principal optical axis by retrieving the generalized linear geometric phase at fundamental frequency, which gives us the relation $\theta_{eff} = 3\theta$ for C3 silicon meta-atoms in square lattice, and Figure 2d schematically shows the rotation of the effective principal axis when varying the local orientation angle $\theta$. Next, eight wave-mixing processes are theoretically allowed according to Eq. (1), which are characterized by the effective third-order nonlinear susceptibility tensor, i.e.: $\chi^{(3)}_{LLLL}$, $\chi^{(3)}_{LLLR}$, $\chi^{(3)}_{LLRR}$, $\chi^{(3)}_{LRRR}$ for the cross-polarized conversion process, which correspond to the phases $\exp(j12\theta)$, $\exp(j6\theta)$, 1, $\exp(-j6\theta)$; $\chi^{(3)}_{RLLL}$, $\chi^{(3)}_{RLLR}$, $\chi^{(3)}_{RLRR}$, $\chi^{(3)}_{RRRR}$ for the co-polarized conversion process, which correspond to the phases $\exp(j6\theta), 1, \exp(-j6\theta), \exp(-j12\theta)$. Therefore, we can compare the simulation data with the above wave-mixing processes and figure out the origin of the observed nonlinear geometric phases. For example, the nonlinear geometric phase calculated from a C3 meta-atom in a square lattice can be understood as the co-polarized conversion process $\chi^{(3)}_{RLLL}$, and the corresponding nonlinear polarization is analytically given as $P_{R'}(3\omega) = \varepsilon_0 \chi^{(3)}_{RLLL} \exp(j6\theta)(E^L(\omega))^3$. Here, $P_{R'}(3\omega)$ refers to the nonlinear polarization contributed to the RCP TH emissions, and $E^L(\omega)$ is the LCP component of the total field excited by the fundamental wave.

Interestingly, when we vary the thickness of the C3 silicon meta-atom to modulate the coupling strength, an optimized height of 775 nm is obtained to unlock the phase modulation capacity of the cross-polarized conversion process (Figure S1). Here, stronger coupling among the meta-atoms can give rise to a nonlinear geometric phase of $\exp(j12\theta)$ on the cross-polarized TH signals. In this case, the wave-mixing process $\chi^{(3)}_{LLLL}$ is thought to play a major role, and its averaged TH conversion efficiency is about $1.38 \times 10^{-8}$. Similarly, for a C5 silicon meta-atom placed in a square lattice, more complex nonlinear geometric phase modulations can be selectively excited by carefully varying the coupling strength among the units, e.g., two distinct nonlinear phase modulations of co-polarized conversion are accessible by carefully varying the thickness of meta-atoms (Figure S2).

*Nonlinear geometric phase obtained with C2/C3/C4 meta-atoms placed in hexagonal lattice*

We further investigate the THG of dielectric structures placed in a hexagonal lattice. Generally, the nonlinear geometric phases of TH emissions from silicon meta-atoms of C2 rotational symmetry are the same as that of the C2 structure placed in a square lattice. Figure 3a shows the TH field, and nonlinear phase modulation carried by the TH signals emitted from C2 structures are proportional to $\exp(j2\theta)$ and $\exp(j4\theta)$, respectively. Here, the operating fundamental wavelength is 1250 nm, the lattice constant is 430 nm, the length and width of the C2 structure are 330 nm and 130 nm. However, for a C3 meta-atom placed in a hexagonal lattice, the TH field is mainly bounded in the vicinity of the silicon meta-atoms and no emission can be observed in the far-field, which is consistent with the selection rule, see Figure 3b. Here, the geometry of the C3 meta-atom is the same as that in a square lattice, the fundamental wavelength is selected as 1200 nm, and the lattice constant is 450 nm.

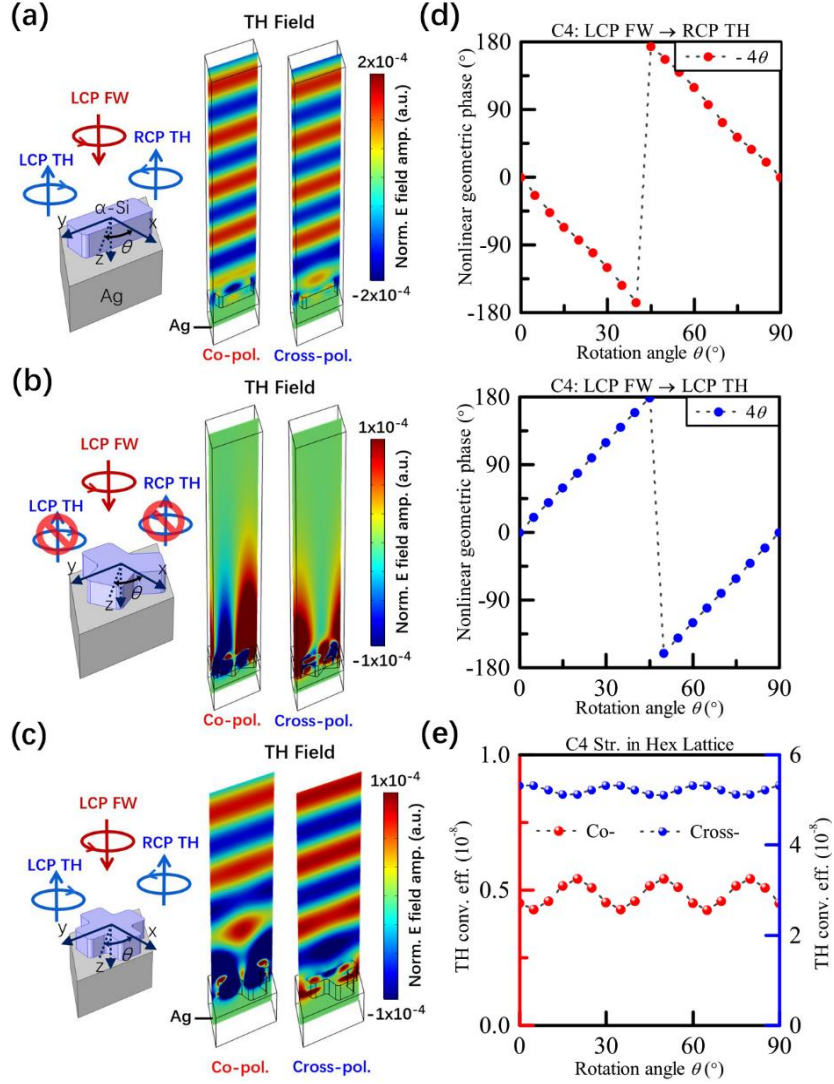

**Figure 3.** THG of meta-atoms placed in a hexagonal lattice. TH fields and the nonlinear geometric phases of (a) C2, (b) C3 and (c) C4 silicon meta-atom. (d) Nonlinear geometric phase of co-polarized and cross-polarized nonlinear conversion processes and (e) corresponding nonlinear conversion efficiency in C4 silicon meta-atom array with varied local orientation angle $\theta$.

Interestingly, a C4 silicon meta-atom placed in a hexagonal lattice also shows a generalized geometric-phase modulation for both co-polarized and cross-polarized TH emissions while the TH conversion efficiency is relatively high, see Figure 3c. Due to the symmetry mismatch between the C4 meta-atom and hexagonal lattice, the co-polarized components of TH signals can radiate to the far-field. According to our simulation data, the nonlinear geometric phase of co-polarized and cross-polarized TH signals are proportional to $\exp(-j4\theta)$ and $\exp(j4\theta)$, respectively, see Figure 3d. The averaged nonlinear conversion efficiency of co-polarized and cross-polarized TH signals are about $4.8\times10^{-9}$ and $5.2\times10^{-8}$, see Figure 3e. Notably, the generalized linear geometric phase of the C4 structure obtained at the fundamental frequency is $\exp(-j4\theta)$, which gives the relation $\theta_{eff} = -2\theta$ (Figure S2). Therefore, the co-polarized and cross-polarized TH emissions can be interpreted as the major contribution from $\chi^{(3)}_{RLLL}$ and $\chi^{(3)}_{LLLL}$, respectively. In addition, for the C5 silicon meta-atom placed in a hexagonal lattice, we also observe generalized nonlinear geometric phases associated with the co-polarized and cross-polarized conversion processes, which are proportional to $\exp(-j10\theta)$ and $\exp(j10\theta)$, respectively (Figure S4).

**Table 1.** Generalized linear geometric phase obtained with silicon meta-atoms with C1-C5 rotational symmetries in the square or hexagonal lattice.

| Lattice type | Structure symmetry type | | | | |
|---|---|---|---|---|---|
| | **C1** | **C2** | **C3** | **C4** | **C5** |
| **Square** | $2\sigma\theta$ | $2\sigma\theta$ | $6\sigma\theta$ | | $10\sigma\theta$ |
| **Hexagonal** | $2\sigma\theta$ | $2\sigma\theta$ | | $-4\sigma\theta$ | $-10\sigma\theta$ |

**Table 2.** Generalized nonlinear geometric phase obtained with silicon meta-atoms with C1-C5 rotational symmetries in the square or hexagonal lattice.

| Lattice type | Pol. | Structure symmetry type | | | | |
|---|---|---|---|---|---|---|
| | | **C1** | **C2** | **C3** | **C4** | **C5** |
| **Square** | **Co-pol.** | $2\sigma\theta$ | $2\sigma\theta$ | $6\sigma\theta$ | | $\pm10\sigma\theta$ |
| | **Cross-pol.** | $4\sigma\theta$ | $4\sigma\theta$ | $12\sigma\theta$ | $4\sigma\theta$ | – |
| **Hexagonal** | **Co-pol.** | $2\sigma\theta$ | $2\sigma\theta$ | | $-4\sigma\theta$ | $-10\sigma\theta$ |
| | **Cross-pol.** | $4\sigma\theta$ | $4\sigma\theta$ | | $4\sigma\theta$ | $10\sigma\theta$ |

As a summary, Table 1 and 2 correspondingly gives the generalized linear and nonlinear geometric phase obtained with different combinations of structural symmetry and lattice symmetry. And our wave-mixing model can well explain the nontrivial nonlinear geometric phases that are retrieved from the simulations.

*Experiments*

In this section, we experimentally investigate the THG of thin dielectric metasurfaces, which includes the generalized nonlinear geometric phase associated with the TH emissions from C2/C3/C4 meta-atoms in square lattice and C2/C4 meta-atoms in hexagonal lattice. Here, phase gradient metasurfaces composed of meta-atoms of gradually varied local orientation angles are fabricated, which can deflect the phase-modulated TH signals to the desired diffraction orders. Figure 4a shows our experimental setup. A pulsed laser of 200 fs pulse width is generated with an optical parametric oscillator and functions as FW light source. Next, the FW passes through a linear polarizer and a quarter waveplate to generate a circular polarization (CP) state. Since dichroic mirror or beam splitter reflections tend to distort CP states due to the different reflection amplitudes and phases for s- and p-polarized components of light, here we use a mirror at a small angle of incidence (AOI) of ~9° to minimize this effect (related discussion can be found in the Supporting Information). The FW is focused onto the sample metasurface (MS) by using a focusing lens slightly off-axis. Because of this, the sample is illuminated at an AOI of ~2° and the generated TH is collimated with a small offset to the incident beam that allows the TH to bypass the mirror. Additionally, we use a telescopic setup to reduce the spatial distance between the diffraction orders to match the size of the camera detector chip and a polarization analyzer to determine the polarization states of the TH diffraction orders. Figure 4b is the scanning electron microscopy (SEM) images of gradient metasurface made of C2/C4 silicon meta-atoms placed in square lattice and C2 silicon meta-atoms placed in hexagonal lattice. Details of the fabrication process can be found in Supporting Information (Figure S7).

Figure 4c shows the TH diffraction pattern of the gradient metasurface made of C2 silicon meta-atom placed in a square lattice whose unit period is 400 nm, and one supercell contains 20 meta-atoms. Here, the $0^{th}$ order, which should locate at the white circle, is removed to improve the clarity of other diffraction orders. In our experiment, the fundamental wavelength is selected as 1240 nm (same for the following three measurements). The measured co-polarized and cross-polarized diffraction angles of the expected diffracted spots are about $-3.1° \pm 0.1°$ and $-5.8° \pm 0.1°$, which are close to the theoretical value of $-2.96°$ and $-5.93°$, respectively.

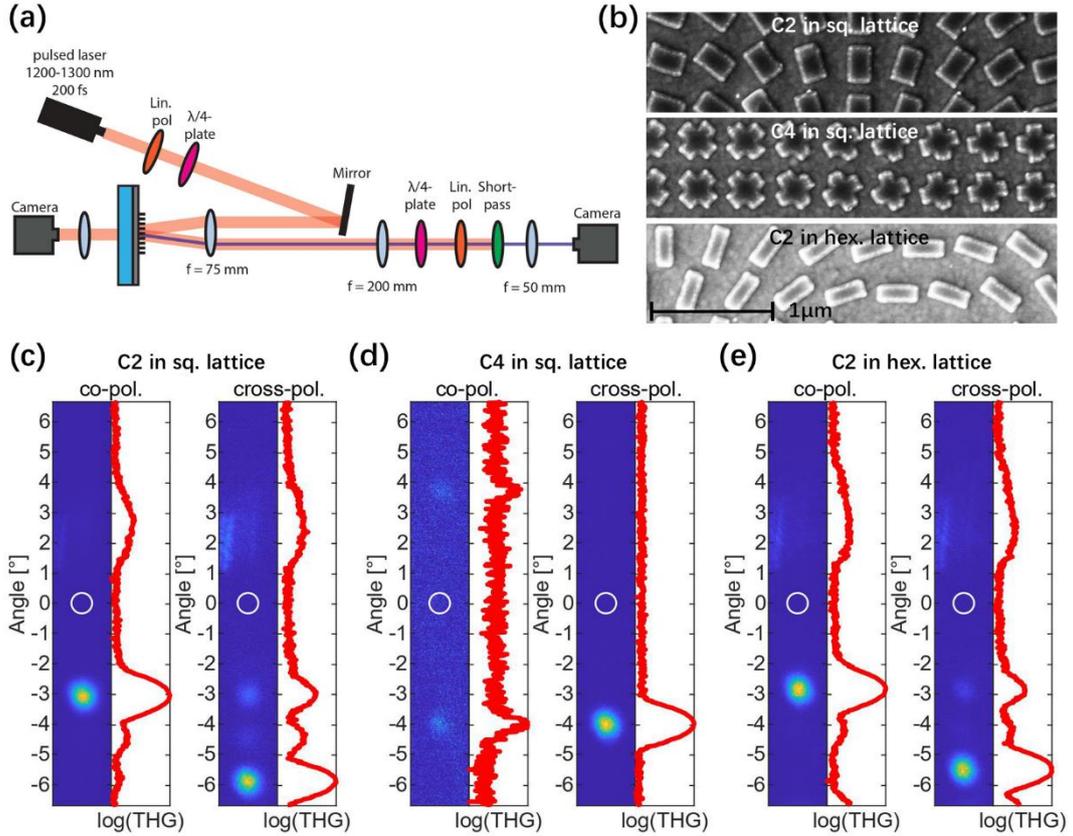

**Figure 4.** Experimental measurements of THG in gradient metasurface made of thin silicon meta-atoms. (a) Experimental setup. (b) SEM image of the fabricated gradient metasurfaces. Measured diffraction pattern of co-polarized and cross-polarized TH emissions from (c) C2, (d) C4 silicon meta-atom placed in a square lattice, and (e) C2 silicon meta-atom placed in a hexagonal lattice.

For the metasurface made of C4 silicon meta-atoms placed in square lattice whose unit period is 400 nm (Figure 4d), one supercell contains 15 meta-atoms. Here, we do not observe apparent diffraction orders for the co-polarized TH signals (except the strong $0^{th}$ order) while cross-polarized TH emissions are deflected to the $+1^{st}$ diffraction order whose diffraction angle is about $-3.9° \pm 0.1°$, which is close to the theoretical value of $-3.95°$. In addition, for the metasurface made of C2 silicon meta-atoms placed in a hexagonal lattice whose unit period is 430 nm (Figure 4e), one supercell contains 20 meta-atoms, and the measured co-polarized and cross-polarized diffraction angles are about –2.8° ± 0.1° and –5.5° ± 0.1°, which are close to the theoretical values of –2.75° and –5.52°, respectively.

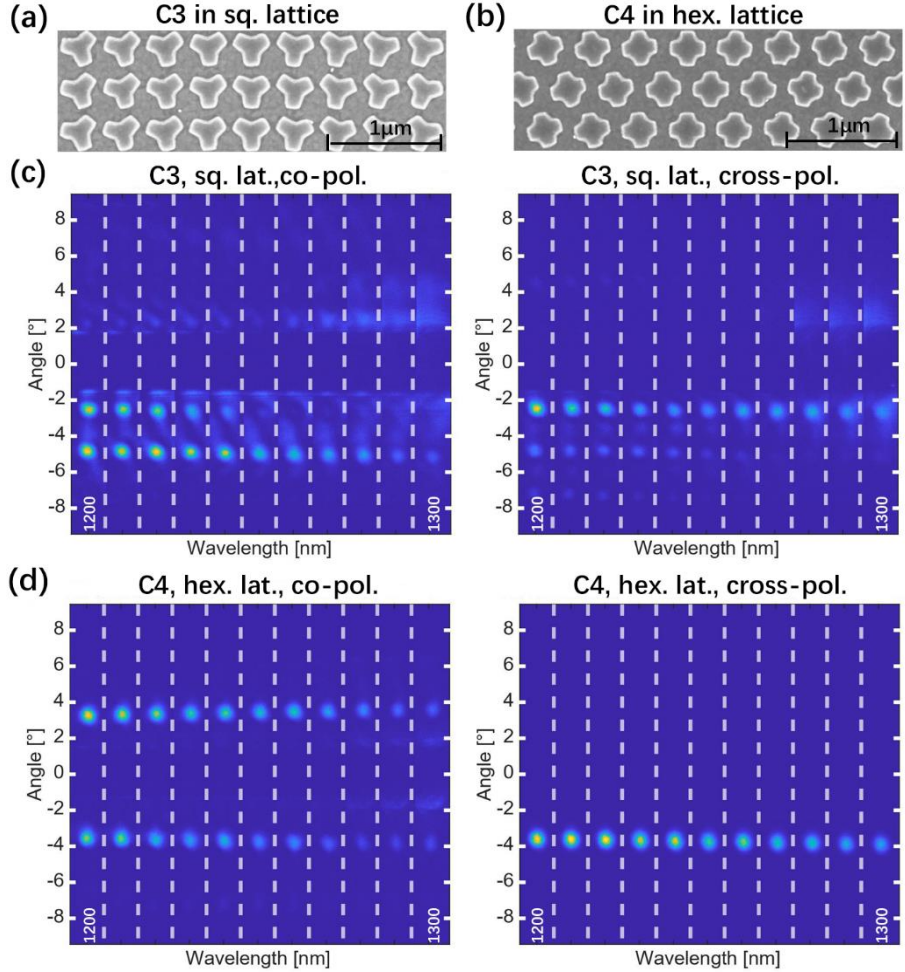

**Figure 5.** Experimental verification of the generalized nonlinear geometric phase in thin silicon meta-atoms. SEM image of the fabricated gradient metasurfaces made of (a) C3 meta-atoms placed in square lattice and (b) C4 meta-atoms placed in a hexagonal lattice. Measured diffraction pattern of the TH emissions of (c) C3 silicon meta-atom in a square lattice and (d) C4 silicon meta-atom when varying the fundamental wavelength from 1200 nm to 1300 nm in steps of 10 nm.

The above measurements generally fit and reflect the influence of nonlinear geometric phases that obey the selection rules. Next, we further experimentally explore the TH emissions from gradient metasurfaces made of C3 silicon meta-atom placed in a square lattice and C4 silicon meta-atom placed in a hexagonal lattice. SEM images of the two samples are shown in Figure 5a,b. In this work, the key evidence of nonlinear geometric-phase modulation on TH emission can be inferred from its asymmetric and spin-dependent diffraction pattern. As shown and discussed in the theoretical part, the mismatch between structure and lattice symmetry makes it possible to couple out the trapped TH energy to far-field, therefore, TH diffraction patterns of the above two samples would be sensitive to the coupling and scattering among meta-atoms.

Figure 5c shows the TH diffraction pattern of the C3 metasurface by varying the fundamental wavelength from 1200 nm to 1300 nm by step of 10 nm, where the strong $0^{th}$-order is removed to improve the clarity of other diffraction orders. Here, the C3 metasurface is composed of the periodic supercell that contains 24 meta-atoms and the unit period is 390 nm. According to the measurement, TH energy is mainly coupled to the $1^{st}$ and $2^{nd}$ diffraction orders. The bright $2^{nd}$ order diffraction spot of the co-polarized TH emissions is attributed to the nonlinear conversion process $\chi^{(3)}_{RLLL} \exp(j6\theta)$. For example, considering the fundamental wavelength of 1200 nm, the measured diffraction angle of the $2^{nd}$ diffraction order is about $-4.9° \pm 0.1°$ and is very close to the theoretical value of $4.90°$, which is consistent with our theory. Other weak diffraction spots can be attributed to the Bragg scattering (especially the cross-polarized emissions). It should be noted that the Bragg scattering in this work refers to the constructive interference of the third-harmonic signals decoupled from the periodic meta-atoms in gradient metasurface. Here, the relatively strong $1^{st}$ order diffractions are contributed by the Bragg scattering from the $0^{th}$ and $2^{nd}$ order, of which the diffracted angle (at 1200 nm) is $-2.5° \pm 0.1°$ and close to the value calculated by the reciprocal lattice vector ($-2.51°$). Another important factor we should be aware of is that the intensity of TH emissions would divert from the expected value obtained with the periodic meta-atoms, which gives nonlinear phase gradient metasurface additional amplitude modulation.

For the phase gradient metasurface made of C4 meta-atoms placed in hexagonal lattice, its supercell contains 15 meta-atoms and unit period is 430 nm, and we observed apparent co-polarized and cross-polarized TH emissions, see Figure 5d. In our measurement, bright $\pm 1^{st}$ order diffractions coexist in co-polarized TH emissions. Considering the fundamental wavelength of 1200 nm, the measured angle of $\pm 1^{st}$ order diffraction is −3.6°±0.1° and 3.4°±0.1°. According to our theory, the ±1st order diffractions are attributed to the co-polarized nonlinear conversion processes $\chi^{(3)}_{RLRR}\exp(j4\theta)$ and $\chi^{(3)}_{RLLL}\exp(-j4\theta)$, respectively. A similar analysis can also be carried out for cross-polarized TH emissions, i.e., the bright 1st order is attributed to the cross-polarized nonlinear conversion processes $\chi^{(3)}_{LLLL}\exp(j4\theta)$, and the measured diffracted angle of −3.6°±0.1° is close to the theoretical value of −3.56°.

It should be noted that the generalized nonlinear geometric phases shown in simulations, especially those that are obtained with the configuration of the C3 meta-atom in a square lattice or the C4 meta-atom in a hexagonal lattice, are indeed retrieved from the periodic meta-atom array. Therefore, the generalized nonlinear geometric phase itself strongly relies on the nonlocal response of the meta-atom array. However, in a real gradient metasurface or other nonlinear meta-devices, the local orientation angles of meta-atoms would vary pixel by pixel and the coupling strength among neighboring units would deviate from the ideal periodic configuration. According to our mode-coupling theory, when the coupling strength is varied, the TH emissions can have contributions by different wave-mixing processes, therefore, the overall TH emissions would contain the signals contributed by several nonlinear conversion processes, which finally gives numerous diffraction orders.

It is worth mentioning that our wave-mixing model can also be applied to explain the nonlinear phase modulation of SHG in our metasurface. Here the second-harmonic (SH) emissions should be contributed by both the surface of silver substrate and silicon meta-atom. The nonlinear currents stimulated on the surface of silver substrate or silicon meta-atom are proportional to the modes excited in dielectric meta-atoms, in this case, the geometric phase of SH emissions can be modeled by an equivalent transverse second-order nonlinear susceptibility. Detailed discussion on the theory and experiments can be found in Supporting Information (Figure S6).

**CONCLUSION**

In conclusion, we theoretically investigate the THG in thin silicon nanostructures of various rotational symmetries and further explore the nonlinear geometric phases of the TH emissions. By replacing the transparent insulator substrate with a reflective metallic substrate, efficient third-harmonic generations are obtained that improve the conversion efficiency by orders of magnitude. In addition, the combination of the structure and lattice rotational symmetry results in numerous types of nonlinear geometric phases that are far beyond the options given by the selection rule. The underlying mechanism can be understood via the strong coupling among the meta-atoms in the dense array resulting in a generalized nonlinear geometric phases description. Our work provides a wave-mixing picture that modifies the linear geometric phase modulation based on different modes with a generalized term, which offers the strategy to predict the possible nonlinear phase terms from their linear response. This work extends the limit of the selection rule and offers new possibilities in nonlinear optical field manipulations.

## ASSOCIATED CONTENT

**Supporting Information**

Supporting Information available: wave-mixing theory for linear and nonlinear geometric phase, full-wave simulation results of other combinations of structure and lattice symmetry, influence of surface plasmons on THG, SHG measurements, flow chart of sample fabrication and experiment setup optimization. Data supporting this study are openly available from Zenodo at DOI 10.5281/zenodo.10043784.

**Author Contributions**

B.L. proposed the concept, developed the theory, and conducted the numerical simulations. R.G. conducted the sample fabrication and implemented nonlinear optical experiments. B.L., R.G., Z.S., K.G., Y.W., Z.G., L.H., T.Z. discussed the results. B.L., R.G., Z.S., L.H., T.Z. wrote the manuscript.

## ACKNOWLEDGMENT

The help offered by Dr. Yuhong Na (U) from Anhui University in manuscript writing, and the discussions with Dr. Wenlong Gao, Dr. Bernhard Reineke Matsudo and Dr. Jinlong Lu from Paderborn University are acknowledged. B.L. acknowledges the financial support by the National Natural Science Foundation of China (Grant No. 12104044). L.H. acknowledges the financial support by the National Natural Science Foundation of China (Grant No. 92050117, U21A20140), National Key R&D Program of China (2021YFA1401200), Beijing Outstanding Young Scientist Program (BJJWZYJH01201910007022), Beijing Municipal Science & Technology Commission, Administrative Commission of Zhongguancun Science Park (Z211100004821009), and Fok Ying-Tong Education Foundation of China (No.161009). T.Z. acknowledges the financial support by the Deutsche Forschungsgemeinschaft (DFG) Collaborative Research Center TRR 142 (Project No. 231447078, Project B09).

Electro-Optics, Technical Digest Series (Optica Publishing Group, 2022), paper FTh1A.7.

# Supplementary Information

## Nonlinear dielectric geometric-phase metasurface with simultaneous structure and lattice symmetry design

Bingyi Liu[1,§], René Geromel[2,*], Zhaoxian Su[3], Kai Guo[1], Yongtian Wang[3], Zhongyi Guo[1], Lingling Huang[3,**], and Thomas Zentgraf[2,4,††]

[1] *School of Computer Science and Information Engineering, Hefei University of Technology, Hefei, 230009, China*

[2] *Department of Physics, Paderborn University, Paderborn, 33098, Germany*

[3] *School of Optics and Photonics, Beijing Engineering Research Center of Mixed Reality and Advanced Display, Beijing Institute of Technology, Beijing, 100081, China*

[4] *Institute for Photonic Quantum Systems, Paderborn University, Paderborn, 33098, Germany*

Supporting Information includes 20 pages, 6 sections and 8 figures, which focuses on: (1) wave-mixing theory for linear and nonlinear geometric phase; (2) full-wave simulation results of other combinations of structure and lattice symmetry; (3) influence of surface plasmons on THG; (4) SHG measurements; (5) flow chart of sample fabrication; (6) discussion of the influence of variation of polarization state of fundamental wave on the TH measurements.

---

[§] These two authors contributed equally to this work.
[**] Email: huanglingling@bit.edu.cn
[††] Email: thomas.zentgraf@uni-paderborn.de

# 1. Interpretation of linear geometric phase via representation transformation

In this section, we show the insight of linear geometric phase from the view of representation transformation, which is a distinct description of the appearance of the linear geometric phase beside the reported Jones calculus-based model. The physical picture of the scattering process occurring in dielectric nanostructures is addressed by the secondary radiations caused by the induced polarizations, which consists of linear term and high-order nonlinear terms.

As illustrated in Figure 1, the forward linear polarization representation (FLPR) is explicitly labeled, and the fundamental wave is normally incident on the meta-atom array along $+z$ direction. In this work, the unit vectors $\mathbf{e}_x$, $\mathbf{e}_y$, $\mathbf{e}_z$ are fixed as the reference to help determine the rotation elements in the forward circular polarization representation (FCPR) and backward circular polarization representation (BCPR). On this basis, the unit vectors $\left(\mathbf{e}_L^\pm, \mathbf{e}_R^\pm, \mathbf{e}_z^\pm\right)$ of FCPR and BCPR are explicitly expressed as:

$$\mathbf{e}_L^\pm = \frac{1}{\sqrt{2}}\left(\mathbf{e}_x \pm j\mathbf{e}_y\right), \mathbf{e}_R^\pm = \frac{1}{\sqrt{2}}\left(\mathbf{e}_x \mp j\mathbf{e}_y\right), \mathbf{e}_z^\pm = \pm\mathbf{e}_z \quad (S1.1)$$

Considering an electric field vector **E** in the above two representations, it can be decomposed into the summation of different components before (without the prime) and after (with the prime) the rotation:

$$\mathbf{E} = \sum_\alpha {}^\pm E^\alpha \mathbf{e}_\alpha^\pm = \sum_{\alpha'} {}^\pm E^{\alpha'} \mathbf{e}_{\alpha'}^\pm \quad (S1.2)$$

Here ${}^\pm E^\alpha$ is the contravariant component of the electric field in FCPR or BCPR, and $\alpha$ refers to the base of circular polarization representation (CPR), i.e., *L*, *R*, and *z*, which corresponds to the values of 1, –1, and 0, respectively. Eq. (S1.2) indicates that the electric field vector **E** is independent of the choice of representation.

In FLPR (exactly a Cartesian coordinate), when the frame that is rotated by an angle $\theta$ around its z-axis gives the following two relations [1]:

$$\begin{bmatrix} \mathbf{e}_{x'} \\ \mathbf{e}_{y'} \\ \mathbf{e}_{z'} \end{bmatrix} = \begin{bmatrix} \cos\theta & \sin\theta & 0 \\ -\sin\theta & \cos\theta & 0 \\ 0 & 0 & 1 \end{bmatrix} \begin{bmatrix} \mathbf{e}_x \\ \mathbf{e}_y \\ \mathbf{e}_z \end{bmatrix}, \quad \begin{bmatrix} E^{x'} \\ E^{y'} \\ E^{z'} \end{bmatrix} = \begin{bmatrix} \cos\theta & -\sin\theta & 0 \\ \sin\theta & \cos\theta & 0 \\ 0 & 0 & 1 \end{bmatrix} \begin{bmatrix} E^x \\ E^y \\ E^z \end{bmatrix} \quad (S1.3)$$

Based on the relation given in Eq. (S1.1), we could further derive the rotation elements in CPR:

$$\begin{bmatrix} \mathbf{e}^{\pm}_{L'} \\ \mathbf{e}^{\pm}_{R'} \\ \mathbf{e}^{\pm}_{z'} \end{bmatrix} = \begin{bmatrix} e^{\pm j\theta} & 0 & 0 \\ 0 & e^{\mp j\theta} & 0 \\ 0 & 0 & 1 \end{bmatrix} \begin{bmatrix} \mathbf{e}^{\pm}_L \\ \mathbf{e}^{\pm}_R \\ \mathbf{e}^{\pm}_z \end{bmatrix}, \quad \begin{bmatrix} {}^{\pm}E^{L'} \\ {}^{\pm}E^{R'} \\ {}^{\pm}E^{z'} \end{bmatrix} = \begin{bmatrix} e^{\mp j\theta} & 0 & 0 \\ 0 & e^{\pm j\theta} & 0 \\ 0 & 0 & 1 \end{bmatrix} \begin{bmatrix} {}^{\pm}E^L \\ {}^{\pm}E^R \\ {}^{\pm}E^z \end{bmatrix} \quad (S1.4)$$

Specifically speaking, for FCPR, we have:

$$\begin{aligned} \mathbf{e}^{+}_{\alpha'} &= {}^{+}R^{\alpha}{}_{\alpha'}(\theta)\mathbf{e}^{+}_{\alpha} = \delta_{\alpha'\alpha}\exp(j\alpha\theta)\mathbf{e}^{+}_{\alpha}, \\ {}^{+}E^{\alpha'} &= {}^{+}R_{\alpha}{}^{\alpha'}(\theta){}^{+}E^{\alpha} = \delta_{\alpha'\alpha}\exp(-j\alpha\theta){}^{+}E^{\alpha} \end{aligned} \quad (S1.5)$$

For BCPR, we have:

$$\begin{aligned} \mathbf{e}^{-}_{\alpha'} &= {}^{-}R^{\alpha}{}_{\alpha'}(\theta)\mathbf{e}^{-}_{\alpha} = \delta_{\alpha'\alpha}\exp(-j\alpha\theta)\mathbf{e}^{-}_{\alpha}, \\ {}^{-}E^{\alpha'} &= {}^{-}R_{\alpha}{}^{\alpha'}(\theta){}^{-}E^{\alpha} = \delta_{\alpha'\alpha}\exp(j\alpha\theta){}^{-}E^{\alpha} \end{aligned} \quad (S1.6)$$

Here, $\delta_{\alpha'\alpha}$ is Kronecker delta, ${}^{\pm}R_{\alpha}{}^{\alpha'}(\theta) = \delta_{\alpha'\alpha}\exp(\mp j\alpha\theta)$ is the elements of rotation operator in FCPR or BCPR, and:

$$\pm R^{\alpha}{}_{\alpha'}(\theta) = \left[ {}^{\pm}R_{\alpha}{}^{\alpha'}(\theta) \right]^{-1} = \exp(\pm j\alpha\theta) \quad (S1.7)$$

Obviously, Eq. (S1.7) is the natural result of Eq. (S1.2).

Next, considering linear polarization $\mathbf{P}(\omega)$ excited in the dielectric meta-atom, which can be decomposed to a similar form that is shown in Eq. (S1.2):

$$\mathbf{P}(\omega) = \sum_{\alpha} {}^{\pm}P^{\alpha}\mathbf{e}^{\pm}_{\alpha} = \sum_{\alpha'} {}^{\pm}P^{\alpha'}\mathbf{e}^{\pm}_{\alpha'} \quad (S1.8)$$

Since incident beams are reflected to the upper half space, we correspond the linear polarization to the reflections in BCPR:

$$\begin{aligned} {}^{-}P^{\alpha'}(\omega) &= \sum_{\alpha} {}^{-}R_{\alpha}{}^{\alpha'}(\theta){}^{-}P^{\alpha}(\omega) \\ &= \varepsilon_0 \sum_{\alpha,\beta} {}^{-}R_{\alpha}{}^{\alpha'}(\theta)\chi^{(1)}_{\alpha\beta}{}^{-}E^{\beta}(\omega) \end{aligned} \quad (S1.9)$$

Here, $\chi^{(1)}_{\alpha\beta}$ is the linear susceptibility of dielectric, ${}^{-}E^{\beta}(\omega)$ is the total field excited in the meta-atom array, and we further have:

$$\begin{aligned}
{}^-E^{\beta}(\omega) &= \sum_{\beta'} {}^-R_{\beta}^{\ \beta'}(\theta)\, {}^-E^{\beta'}(\omega) \\
&= \sum_{\beta'} \left[ {}^+R_{\beta}^{\ \beta'}(\theta) \right]^{-1} {}^-E^{\beta'}(\omega)
\end{aligned} \quad (S1.10)$$

Combing Eq. (S1.9) and (S1.10), we have:

$${}^-P^{\alpha'}(\omega) = \varepsilon_0 \sum_{\alpha,\beta} \sum_{\beta'} {}^-R_{\alpha}^{\ \alpha'}(\theta)\, \chi_{\alpha\beta}^{(1)} \left[ {}^+R_{\beta}^{\ \beta'}(\theta) \right]^{-1} {}^-E^{\beta'}(\omega) \quad (S1.11)$$

Because

$${}^-P^{\alpha'}(\omega) = \varepsilon_0 \sum_{\beta'} \chi_{\alpha'\beta'}^{(1)}\, {}^-E^{\beta'}(\omega) \quad (S1.12)$$

Then we have:

$$\chi_{\alpha'\beta'}^{(1)} = \sum_{\alpha,\beta} {}^-R_{\alpha}^{\ \alpha'}(\theta)\, \chi_{\alpha\beta}^{(1)} \left[ {}^+R_{\beta}^{\ \beta'}(\theta) \right]^{-1} \quad (S1.13)$$

Therefore, the co-polarization conversion process defined in reflection-type geometric metasurface corresponds to the LCP (RCP) incidence and RCP (LCP) reflection, which gives:

$$\begin{aligned}
\chi_{R'L'}^{(1)} &= {}^-R_R^{\ R'}(\theta)\, \chi_{RL}^{(1)} \left[ {}^+R_L^{\ L'}(\theta) \right]^{-1} = \chi_{RL}^{(1)} \\
\chi_{L'R'}^{(1)} &= {}^-R_L^{\ L'}(\theta)\, \chi_{LR}^{(1)} \left[ {}^+R_R^{\ R'}(\theta) \right]^{-1} = \chi_{LR}^{(1)}
\end{aligned} \quad (S1.14)$$

Meanwhile, the cross-polarization conversion process defined in reflection-type geometric metasurface corresponds to the LCP (RCP) incidence and LCP (RCP) reflection, which gives:

$$\begin{aligned}
\chi_{L'L'}^{(1)} &= {}^-R_L^{\ L'}(\theta)\, \chi_{LL}^{(1)} \left[ {}^+R_L^{\ L'}(\theta) \right]^{-1} = \chi_{LL}^{(1)} \exp(j2\theta) \\
\chi_{R'R'}^{(1)} &= {}^-R_R^{\ R'}(\theta)\, \chi_{RR}^{(1)} \left[ {}^+R_R^{\ R'}(\theta) \right]^{-1} = \chi_{RR}^{(1)} \exp(-j2\theta)
\end{aligned} \quad (S1.15)$$

Here, we assume that the coupling among the nanostructures is weak, then the rotation angle of the local orientation angle of the structure is the same as the rotation angle of the effective principal optical axis of the meta-atom array, i.e., $\theta_{eff} = \theta$.

Generally speaking, the above two angles do not need to be the same, where the coupling among the meta-atoms would deviate the modulation of the effective principal optical axis from the primary axis of symmetry of the structure, and it has been revealed to obey a more generalized relation, i.e., $\theta_{eff} = l\theta$, $l$ is an integer [3]. Then we redefine the geometric phase term in Eq. (S1.15) as $\exp(-j\Phi)$, and the integer *l* can be simply calculated by $l = \Phi/2\theta$. It should be noted that this step is necessary in our study to determine the elements of rotation operators, which are the basics for the calculation of nonlinear geometric phases.

## 2. Wave-mixing model and an accompanying algebra for the generalized nonlinear geometric phase

In this section, we move forward to the derivation of nonlinear geometric phases of reflection-type metasurface via the wave-mixing model and representation transformation.

Similar to Eq. (S1.9), the third-order nonlinear polarization in dielectric meta-atoms is analytically given as:

$$
\begin{aligned}
{}^-P^{\alpha'}(3\omega) &= \sum_{\alpha} {}^-R_\alpha^{\alpha'} \, {}^-P^\alpha(3\omega) \\
&= \varepsilon_0 \sum_{\alpha,\beta,\gamma,\delta} {}^-R_\alpha^{\alpha'} \chi^{(3)}_{\alpha\beta\gamma\delta} \, {}^-E^\beta(\omega) \, {}^-E^\gamma(\omega) \, {}^-E^\delta(\omega) \\
&= \varepsilon_0 \sum_{\alpha,\beta,\gamma,\delta} \sum_{\beta',\gamma',\delta'} \left( \begin{array}{c} {}^-R_\alpha^{\alpha'} \chi^{(3)}_{\alpha\beta\gamma\delta} \left[{}^+R_\beta^{\beta'}\right]^{-1} \left[{}^+R_\gamma^{\gamma'}\right]^{-1} \left[{}^+R_\delta^{\delta'}\right]^{-1} \bullet \\ {}^-E^{\beta'}(\omega) \, {}^-E^{\gamma'}(\omega) \, {}^-E^{\delta'}(\omega) \end{array} \right)
\end{aligned}
\quad (S2.1)
$$

Here, $\chi^{(3)}_{\alpha\beta\gamma\delta}$ is the third-order nonlinear susceptibility, *α*, *β*, *γ*, and *δ* refer to the base of CPR, i.e., *L*, *R*, and *z*.

Because

$$
{}^-P^{\alpha'}(3\omega) = \varepsilon_0 \sum_{\beta',\gamma',\delta'} \chi^{(3)}_{\alpha'\beta'\gamma'\delta'} \, {}^-E^{\beta'}(\omega) \, {}^-E^{\gamma'}(\omega) \, {}^-E^{\delta'}(\omega) \quad (S2.2)
$$

Then we have:

$$
\chi^{(3)}_{\alpha'\beta'\gamma'\delta'} = \sum_{\alpha,\beta,\gamma,\delta} {}^-R_\alpha^{\alpha'} \chi^{(3)}_{\alpha\beta\gamma\delta} \left[{}^+R_\beta^{\beta'}\right]^{-1} \left[{}^+R_\gamma^{\gamma'}\right]^{-1} \left[{}^+R_\delta^{\delta'}\right]^{-1} \quad (S2.3)
$$

Here, ${}^\pm R_\alpha^{\alpha'} = \delta_{\alpha'\alpha} \exp(\mp j\alpha\theta_{eff})$ is the element of the rotation operator in FCPR and BCPR. Therefore, considering the wave-mixing process between the LCP and RCP components of the modes excited by an LCP fundamental wave, we have eight nonlinear conversion processes.

For the co-polarized nonlinear conversion process, we have:

$$\chi^{(3)}_{R'L'L'L'} = {}^{-}R_R^{R'} \chi^{(3)}_{RLLL} \left[{}^{+}R_L^{L'}\right]^{-1} \left[{}^{+}R_L^{L'}\right]^{-1} \left[{}^{+}R_L^{L'}\right]^{-1} = \chi^{(3)}_{RLLL} \exp(j2\theta_{\text{eff}}) \quad \text{(S2.4a)}$$

$$\chi^{(3)}_{R'L'L'R'} = {}^{-}R_R^{R'} \chi^{(3)}_{RLLR} \left[{}^{+}R_L^{L'}\right]^{-1} \left[{}^{+}R_L^{L'}\right]^{-1} \left[{}^{+}R_R^{R'}\right]^{-1} = \chi^{(3)}_{RLLR} \quad \text{(S2.4b)}$$

$$\chi^{(3)}_{R'L'R'R'} = {}^{-}R_R^{R'} \chi^{(3)}_{RLRR} \left[{}^{+}R_L^{L'}\right]^{-1} \left[{}^{+}R_R^{R'}\right]^{-1} \left[{}^{+}R_R^{R'}\right]^{-1} = \chi^{(3)}_{RLRR} \exp(-j2\theta_{\text{eff}}) \quad \text{(S2.4c)}$$

$$\chi^{(3)}_{R'R'R'R'} = {}^{-}R_R^{R'} \chi^{(3)}_{RRRR} \left[{}^{+}R_R^{R'}\right]^{-1} \left[{}^{+}R_R^{R'}\right]^{-1} \left[{}^{+}R_R^{R'}\right]^{-1} = \chi^{(3)}_{RRRR} \exp(-j4\theta_{\text{eff}}) \quad \text{(S2.4d)}$$

For the cross-polarized nonlinear conversion process, we have:

$$\chi^{(3)}_{L'L'L'L'} = {}^{-}R_L^{L'} \chi^{(3)}_{LLLL} \left[{}^{+}R_L^{L'}\right]^{-1} \left[{}^{+}R_L^{L'}\right]^{-1} \left[{}^{+}R_L^{L'}\right]^{-1} = \chi^{(3)}_{LLLL} \exp(j4\theta_{\text{eff}}) \quad \text{(S2.5a)}$$

$$\chi^{(3)}_{L'L'L'R'} = {}^{-}R_L^{L'} \chi^{(3)}_{LLLR} \left[{}^{+}R_L^{L'}\right]^{-1} \left[{}^{+}R_L^{L'}\right]^{-1} \left[{}^{+}R_R^{R'}\right]^{-1} = \chi^{(3)}_{LLLR} \exp(j2\theta_{\text{eff}}) \quad \text{(S2.5b)}$$

$$\chi^{(3)}_{L'L'R'R'} = {}^{-}R_L^{L'} \chi^{(3)}_{LLRR} \left[{}^{+}R_L^{L'}\right]^{-1} \left[{}^{+}R_R^{R'}\right]^{-1} \left[{}^{+}R_R^{R'}\right]^{-1} = \chi^{(3)}_{LLRR} \quad \text{(S2.5c)}$$

$$\chi^{(3)}_{L'R'R'R'} = {}^{-}R_L^{L'} \chi^{(3)}_{LRRR} \left[{}^{+}R_R^{R'}\right]^{-1} \left[{}^{+}R_R^{R'}\right]^{-1} \left[{}^{+}R_R^{R'}\right]^{-1} = \chi^{(3)}_{LRRR} \exp(-j2\theta_{\text{eff}}) \quad \text{(S2.5d)}$$

According to the relation $\theta_{\text{eff}} = l\theta$, we could obtain the generalized nonlinear geometric phases associated with different wave-mixing processes. Moreover, for the C1/C2 meta-atom in square/hexagonal lattice, $l=1$, in which case the nonlinear geometric phases agree with the phase predicted by the selection rule. However, for the C3/C5 meta-atom in square lattice and the C4/C5 meta-atom in hexagonal lattice, structure symmetry does not match the lattice symmetry, in which case the TH energy generated in silicon could be coupled out to the far-field. Therefore, the dense meta-atom array gives rise to a strong coupling strength among the meta-atoms, which makes the relation between $\theta_{\text{eff}}$ and $\theta$ more complicated.

**(1) C3 meta-atom in square lattice**

For example, consider the C3 meta-atom placed in a square lattice, whose unit thickness is selected as 105 nm. Assuming the LCP fundamental wave incidence, the linear geometric phase associated with the cross-polarized conversion process at 1200 nm is shown in Figure S1(a), which is proportional to $\exp(j6\theta)$, therefore, $\theta_{\text{eff}} = 3\theta$. In this case, all possible nonlinear conversion processes and the corresponding geometric phases can be calculated according to Eq. (S2.4) and (S2.5).

For the co-polarized TH conversion process, we have:

$$\chi^{(3)}_{R'L'L'L'} = \chi^{(3)}_{RLLL} \exp(j6\theta),\ \chi^{(3)}_{R'L'L'R'} = \chi^{(3)}_{RLLR},$$
$$\chi^{(3)}_{R'L'R'R'} = \chi^{(3)}_{RLRR} \exp(-j6\theta),\ \chi^{(3)}_{R'R'R'R'} = \chi^{(3)}_{RRRR} \exp(-j12\theta) \quad (S2.6)$$

For the cross-polarized TH conversion process, we have:

$$\chi^{(3)}_{L'L'L'L'} = \chi^{(3)}_{LLLL} \exp(j12\theta),\ \chi^{(3)}_{L'L'L'R'} = \chi^{(3)}_{LLLR} \exp(j6\theta),$$
$$\chi^{(3)}_{L'L'R'R'} = \chi^{(3)}_{LLRR},\ \chi^{(3)}_{L'R'R'R'} = \chi^{(3)}_{LRRR} \exp(-j6\theta) \quad (S2.7)$$

It should be noted that we generally neglect the wave-mixing processes that involve the z component of the fundamental modes, because the thickness we select for the meta-atom is much smaller than the operating fundamental wavelength, i.e., 1/12 of the fundamental wavelength. However, if the contribution of the wave-mixing processes that involve the z components of the fundamental electric field is considered, some nonlinear phase terms like $\exp(\pm j\theta_{eff})$ would appear.

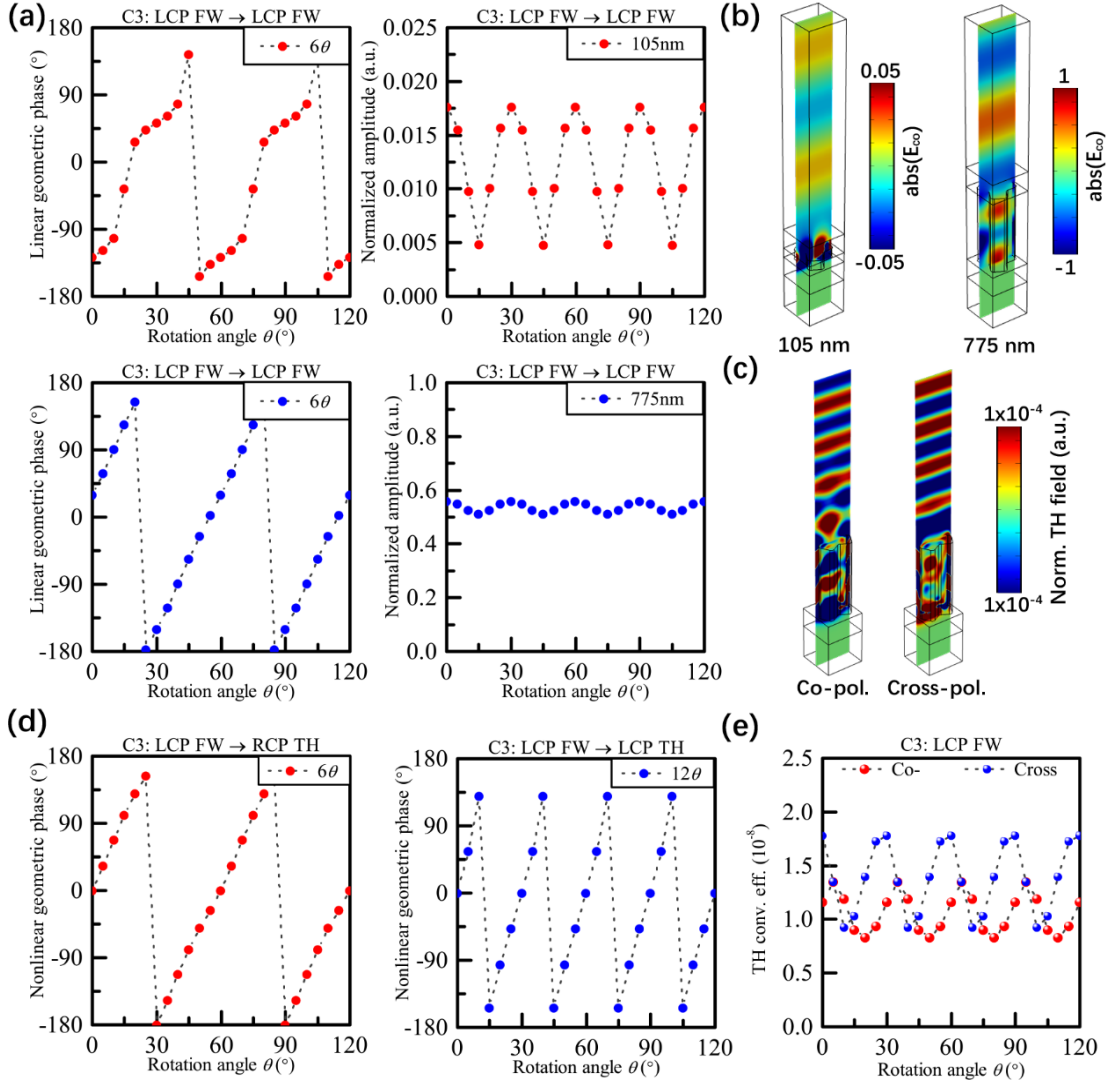

**Figure S1.** (a) Linear geometric phase and the corresponding amplitude of the cross-polarized conversion in C3 silicon meta-atom placed in a square lattice, whose thickness is 105 nm (red dot) and 775 nm (blue dot). (b) Cross-polarized components of the fundamental electric field, whose thickness is 105 nm and 775 nm. (c) TH fields of C3 silicon meta-atom whose thickness is 775 nm. (d) Nonlinear geometric phases and (e) TH conversion efficiency of the co-polarized and cross-polarized conversions.

As discussed in the paper, when we increase the thickness of the C3 meta-atom to enhance the coupling strength, we could obtain the nonlinear geometric phase of cross-polarized conversion which covers the $2\pi$ range. Here, an optimized thickness of 775 nm is obtained. Figure S1(a) shows the linear geometric phase and the corresponding amplitude of the linear cross-polarization conversion in the C3 meta-atom whose thickness is 105 nm and 775 nm, which are both proportional to $\exp(j6\theta)$. Figure S1(b) shows the fundamental field of the cross-polarized electric field. Figure S1(c) shows the co-polarized and cross-polarized TH field of the C3 meta-atom array whose thickness is 775 nm. Figure S1(d) shows the nonlinear geometric phase associated with co-polarized (LCP FW to RCP TH) and cross-polarized (LCP FW to LCP TH) conversions, which correspond to the phase modulation of $\exp(j6\theta)$ and $\exp(j12\theta)$, respectively. Figure S1(e) shows the corresponding TH conversion efficiencies, which are about $1.1\times10^{-8}$ and $1.3\times10^{-8}$.

**(2) C5 meta-atom in square lattice**

As additional discussion on the wave-mixing model characterized by Eq. (1), we could find more evidence in THG of the C5 silicon meta-atom placed in a square lattice. Figure S2(a) shows the schematic of the C5 silicon meta-atom, whose operating wavelength, distance between the vertex and center point, thickness, and period are 1200 nm, 180 nm, 105 nm, and 380 nm, respectively. The far-field co-polarized TH signal shows the geometric phase of $\exp(j10\theta)$ while no phase modulation is observed from the cross-polarized TH signals, see Figure S2(b). The averaged nonlinear conversion efficiency of the above C5 structure is $5\times10^{-12}$, which is relatively low.

However, we can observe a higher TH conversion efficiency by optimizing its thickness, i.e., 510 nm. In this case, the averaged nonlinear conversion efficiency of co-polarized TH signals is increased to $7.6 \times 10^{-8}$, which is improved by 4 orders of magnitude, see Figure S2(d). The geometric phase associated with the co-polarized TH signal is roughly proportional to $\exp(j10\theta)$, but the linearity between the phase and local orientation angle $\theta$ becomes worse, see Figure S2(c). Figure S2(e) shows the generalized linear geometric phase retrieved at fundamental frequency, which is proportional to $\exp(j10\theta)$, then we have $\theta_{eff} = 5\theta$. In addition, the generalized linear geometric phases obtained with C5 meta-tom whose thickness is 105 nm, 500 nm, and 510 nm are all roughly proportional to $\exp(j10\theta)$.

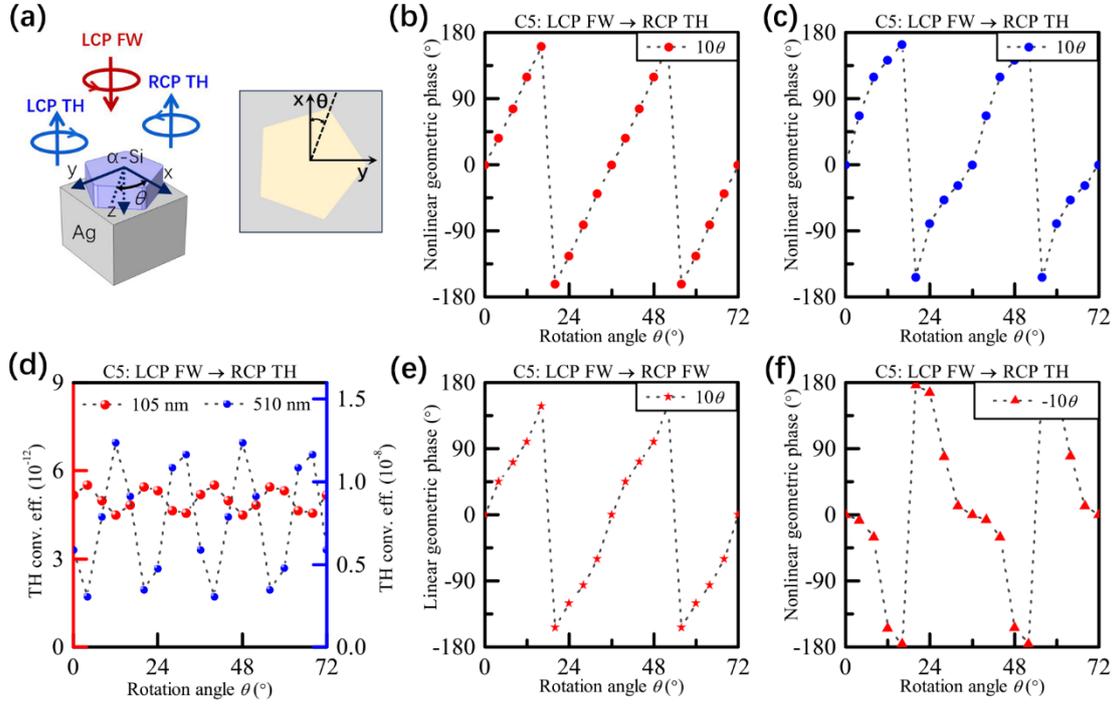

**Figure S2.** (a) Schematic of C5 silicon meta-atom array placed in a square lattice. Nonlinear geometric phase of the co-polarized conversion in meta-atom whose thickness is (b) 105 nm (red dot) and (c) 510 nm (blue dot), and (d) the corresponding nonlinear conversion efficiency. (e) Linear geometric phase of the cross-polarized conversion in meta-atom whose thickness is 510 nm (red star). (f) Nonlinear geometric phase of the co-polarized conversion in meta-atom whose thickness is 500 nm (red rectangle).

According to our wave-mixing model and the rotation elements from the generalized linear geometric phase retrieved at the fundamental frequency, all possible nonlinear conversion processes and the corresponding geometric phases can be calculated. For the co-polarized TH conversion process, we have:

$$\chi^{(3)}_{R'L'L'L'} = \chi^{(3)}_{RLLL}\exp(j10\theta),\ \chi^{(3)}_{R'L'L'R'} = \chi^{(3)}_{RLLR},$$
$$\chi^{(3)}_{R'L'R'R'} = \chi^{(3)}_{RLRR}\exp(-j10\theta),\ \chi^{(3)}_{R'R'R'R'} = \chi^{(3)}_{RRRR}\exp(-j20\theta)$$
(S2.8)

For the cross-polarized TH conversion process, we have:

$$\chi^{(3)}_{L'L'L'L'} = \chi^{(3)}_{LLLL}\exp(j20\theta),\ \chi^{(3)}_{L'L'L'R'} = \chi^{(3)}_{LLLR}\exp(j10\theta),$$
$$\chi^{(3)}_{L'L'R'R'} = \chi^{(3)}_{LLRR},\ \chi^{(3)}_{L'R'R'R'} = \chi^{(3)}_{LRRR}\exp(-j10\theta)$$
(S2.9)

Therefore, the origin of the above nonlinear geometric phase shown in Figure S2(b, c) can be interpreted as the TH emission contributed by the co-polarized conversion process $\chi^{(3)}_{R'L'L'L'} = \chi^{(3)}_{RLLL}\exp(j10\theta)$.

Interestingly, by tuning the thickness of the C5 silicon meta-atom to 500 nm, we observe a conjugate nonlinear geometric phase, i.e., $\exp(-j10\theta)$, see Figure S2(f). Here, the co-polarized conversion process $\chi^{(3)}_{R'L'R'R'} = \chi^{(3)}_{RLRR}\exp(-j10\theta)$ is believed to play a major role. This indicates that when the coupling strength is varied, the wave-mixing process could also be switched.

### (3) C4 meta-atom in hexagonal lattice

In our theory, the nonlinear geometric phase can be calculated with the linear geometric phase retrieved at the fundamental frequency. Figure S3 shows the linear geometric phase retrieved from the C4 silicon meta-atom placed in a hexagonal lattice, which is proportional to $\exp(-j4\theta)$, and we have $\theta_{\text{eff}} = -2\theta$.

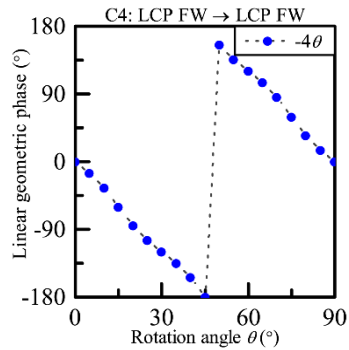

**Figure S3.** Linear geometric phase of the cross-polarized conversion in C4 silicon meta-atom placed in square lattice, whose thickness is 105 nm.

Therefore, all possible nonlinear conversion processes and the corresponding geometric phases can be calculated. For the co-polarized TH conversion process, we have:

$$\chi^{(3)}_{R'L'L'L'} = \chi^{(3)}_{RLLL}\exp(-j4\theta),\ \chi^{(3)}_{R'L'L'R'} = \chi^{(3)}_{RLLR},$$
$$\chi^{(3)}_{R'L'R'R'} = \chi^{(3)}_{RLRR}\exp(j4\theta),\ \chi^{(3)}_{R'R'R'R'} = \chi^{(3)}_{RRRR}\exp(j8\theta)$$
(S2.8)

For the cross-polarized TH conversion process, we have:

$$\chi^{(3)}_{L'L'L'L'} = \chi^{(3)}_{LLLL}\exp(-j8\theta),\ \chi^{(3)}_{L'L'L'R'} = \chi^{(3)}_{LLLR}\exp(-j4\theta),$$
$$\chi^{(3)}_{L'L'R'R'} = \chi^{(3)}_{LLRR},\ \chi^{(3)}_{L'R'R'R'} = \chi^{(3)}_{LRRR}\exp(j4\theta)$$
(S2.9)

### (4) C5 meta-atom in hexagonal lattice

C5 meta-atom placed in a hexagonal lattice (same cross-section as that placed in a square lattice) also shows good nonlinear geometric-phase modulation In our study, the optimized thickness for efficient THG in C5 meta-atom is 350 nm, and the geometric phase carried by the co-polarized and cross-polarized TH waves are proportional to $\exp(-j10\theta)$ and $\exp(j10\theta)$, respectively, see Figure S4(a, b). However, the nonlinear conversion efficiency is reduced by 3 orders of magnitude when compared with the aforementioned demonstrations, see Figure S4(c).

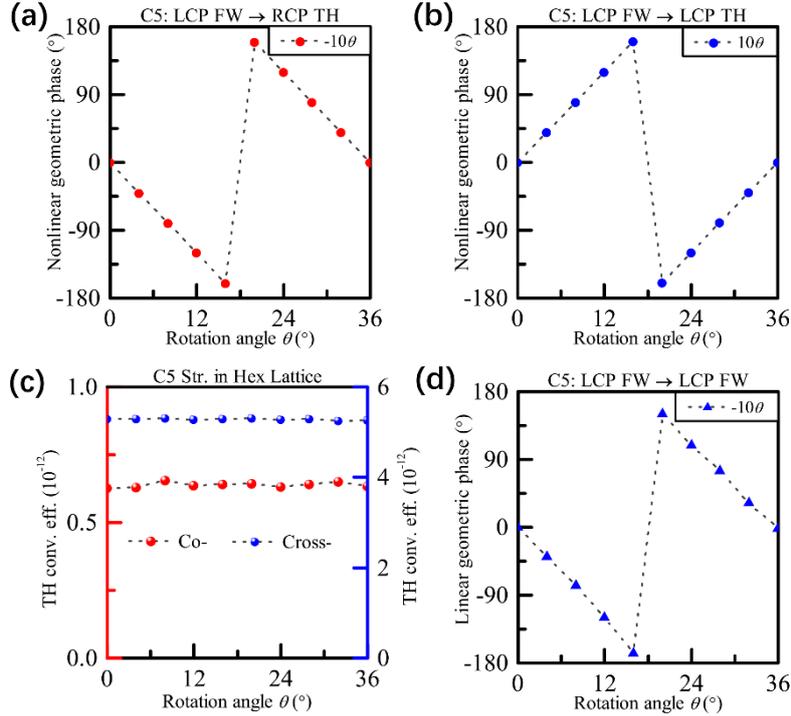

**Figure S4.** Nonlinear geometric phase of (a) the co-polarized and (b) cross-polarized conversion in C5 silicon meta-atom placed in a hexagonal lattice, whose thickness is 350 nm. (c) The corresponding nonlinear conversion efficiency. (d) Linear geometric phase.

The generalized linear geometric phase obtained at the fundamental frequency is $\exp(-j10\theta)$, therefore we have the relation $\theta_{eff} = -5\theta$, see Figure S4(d). In this case, the above nonlinear geometric phases are understood as the contribution of wave-mixing process $\chi^{(3)}_{R'L'L'L'} = \chi^{(3)}_{RLLL}\exp(-j10\theta)$ and $\chi^{(3)}_{L'L'L'L'} = \chi^{(3)}_{LLLL}\exp(j10\theta)$.

In addition, all possible nonlinear conversion processes and the corresponding geometric phases can be calculated. For the co-polarized TH conversion process, we have:

$$\chi^{(3)}_{R'L'L'L'} = \chi^{(3)}_{RLLL}\exp(-j10\theta), \ \chi^{(3)}_{R'L'L'R'} = \chi^{(3)}_{RLLR},$$
$$\chi^{(3)}_{R'L'R'R'} = \chi^{(3)}_{RLRR}\exp(j10\theta), \ \chi^{(3)}_{R'R'R'R'} = \chi^{(3)}_{RRRR}\exp(j20\theta) \quad (S2.10)$$

For the cross-polarized TH conversion process, we have:

$$\chi^{(3)}_{L'L'L'L'} = \chi^{(3)}_{LLLL}\exp(-j20\theta), \ \chi^{(3)}_{L'L'L'R'} = \chi^{(3)}_{LLLR}\exp(-j10\theta),$$
$$\chi^{(3)}_{L'L'R'R'} = \chi^{(3)}_{LLRR}, \ \chi^{(3)}_{L'R'R'R'} = \chi^{(3)}_{LRRR}\exp(j10\theta) \quad (S2.11)$$

Next, we provide a summary of the generalized linear and nonlinear geometric phase modulation that is obtained with the silicon meta-atom of different rotational symmetry and lattice symmetry, see Table S1 and S2. It should be noted that the generalized linear geometric phase we retrieved at the fundamental frequency shows some difference when compared with the table given in Ref [3]. This can be attributed to the different coupling strengths for metallic and dielectric meta-atoms. The generalized linear geometric phase is substantially determined by the modified rotation of the effective principal optical axis, however, it is difficult to analytically give the relation between the modes excited in meta-atoms and the corresponding optical anisotropy (i.e., effective principal optical axis), and one can only evaluate the optical anisotropy of meta-atom array with numerical calculations.

### 3. Influence of surface plasmons on the THG of our metasurfaces

In this section, we discuss the contribution of surface plasmon in THG of our metasurfaces. The surface plasmon channel is an important factor that cannot be ignored for the energy transfer among the dense particles. This can be understood through two aspects: (i) At the fundamental frequency, the surface plasmons would change the coupling intensity among the dense particles, thereby modifying the generalized linear geometric phase; (ii) At the third harmonic frequency, the harmonic signals could also be influenced by coupling the energy to the surface plasmons channel, which would alter the scattering of the third-harmonic signals as well. However, the specific contribution of the surface plasmon in third-harmonic generation and its influence on the coupling among dense dielectric particles is still not clear. In our simulations, the third-harmonic signals generated from the Ag film is considered by implementing nonlinear polarizations of Ag film, however, its contribution is negligible. Interestingly, although the THG in Ag film is negligible, the surface plasmons of the Ag film would efficiently boost the trapping of the fundamental modes in metasurfaces and finally improve the nonlinear conversion efficiency of metasurfaces. This point can be proved by replacing the Ag film with a perfect electric conductor (PEC) boundary, where no surface plasmons can exist.

In this work, the Ag film is modeled with real material, therefore, the participation of possible surface plasmon is naturally considered in our full-wave simulations. Figure S5(a) shows the zoom-in plot of the electric field of C3 silicon meta-atom placed in the square array (region circled by the red dashed rectangle). In our simulation, we consider Ag film as a real material, and the illuminating fundamental wavelength is 1200 nm. Here, we could observe surface plasmons-like field distribution around the Ag film surface. Figure S5(b) is the corresponding third harmonic field when considering the THG in Ag film (region circled by the blue and yellow dashed rectangle), and the third-order nonlinear susceptibility of Ag is $9.3\times10^{-20}$ $m^2/V^2$ in our simulation.

Figure S5(c) shows the fundamental (left) and third-harmonic electric field (right) when replacing Ag with Au, here the third-order nonlinear susceptibility of Au is $8\times10^{-19} m^2/V^2$. Figures S5(d) and (e) are the corresponding TH conversion efficiency and nonlinear geometric phase. The averaged TH conversion efficiency is $2.22\times10^{-9}$ and $3.22\times10^{-10}$ for the co-polarized and cross-polarized conversion process, respectively, which is smaller than the sample supported by Ag substrate. Such difference is attributed to the SPPs excitation in Ag and Au film, which directly influences the trapping of the fundamental wave and thereby changes the intensity of TH signals.

Figure S5(f) shows the fundamental and third-harmonic electric field by replacing the Ag film with the PEC surface, where no surface plasmon resonance could exist.

Figure S5(g) and (h) are the corresponding TH conversion efficiency and nonlinear geometric phases of different polarization conversion processes. The averaged TH conversion efficiency of co-polarized and cross-polarized nonlinear conversion processes are $5.05\times10^{-11}$ and $1.35\times10^{-11}$, respectively. The intensity of the third-harmonic signals is apparently decreased, which indicates the importance of surface plasmons in boosting the THG in metasurfaces.

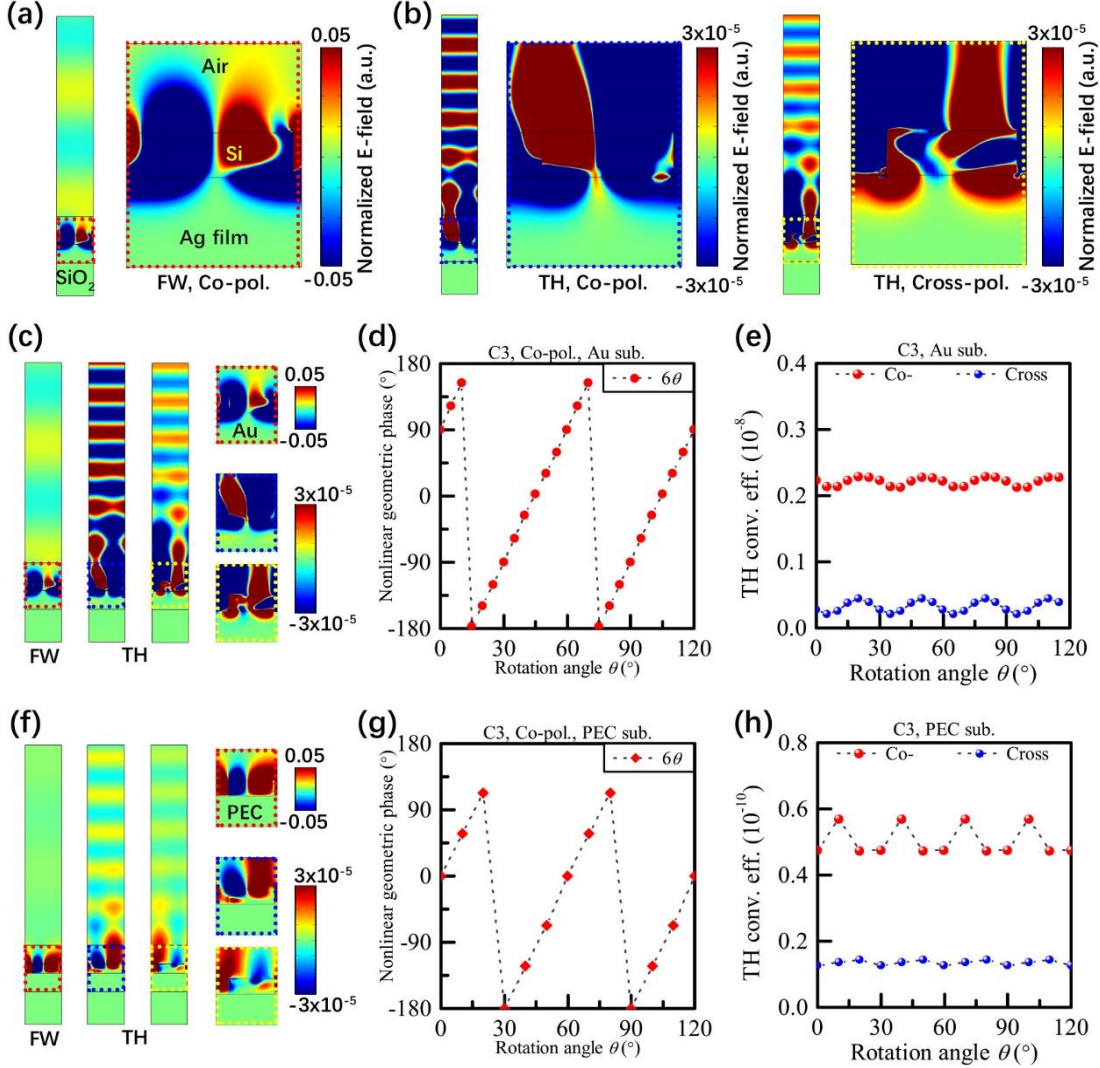

**Figure S5.** (a)Fundamental and (b)TH electric field Ag-substrate metasurface. (c)Fundamental and TH electric field of Au-substrate metasurface. (d)TH conversion efficiency and (e)nonlinear geometric phase of Au-substrate metasurface when varying the local orientation angle. (f)FW and TH electric field of PEC-substrate metasurface. (g)TH conversion efficiency and (h)nonlinear geometric phase of PEC-substrate metasurface when varying the local orientation angle.

## 4. Second harmonic generations in silicon meta-atom placed on a silver substrate

In this section, we discuss the nonlinear geometric phase modulation on the second harmonic (SH). Similar to the third-order nonlinear polarization, we could extend the wave-mixing model given in Section 2 to second-order nonlinear polarization, which gives:

$$\begin{aligned}
{}^-P^{\alpha'}(2\omega) &= \sum_\alpha {}^-R_\alpha^{\alpha'} {}^-P^\alpha(2\omega) \\
&= \varepsilon_0 \sum_{\alpha,\beta,\gamma} {}^-R_\alpha^{\alpha'} \chi^{(2)}_{\alpha\beta\gamma} {}^-E^\beta(\omega) {}^-E^\gamma(\omega) \\
&= \varepsilon_0 \sum_{\alpha,\beta,\gamma} \sum_{\beta',\gamma'} \left( \begin{array}{c} {}^-R_\alpha^{\alpha'} \chi^{(2)}_{\alpha\beta\gamma} \left[{}^+R_\beta^{\beta'}\right]^{-1} \left[{}^+R_\gamma^{\gamma'}\right]^{-1} \cdot \\ {}^-E^{\beta'}(\omega) {}^-E^{\gamma'}(\omega) \end{array} \right)
\end{aligned} \quad (S3.1)$$

Here, $\chi^{(2)}_{\alpha\beta\gamma}$ is the second-order nonlinear susceptibility, α, β, and γ refer to the base of CPR, i.e., $L$, $R$, and $z$.

Because

$$^-P^{\alpha'}(2\omega) = \varepsilon_0 \sum_{\beta',\gamma'} \chi^{(2)}_{\alpha'\beta'\gamma'} {}^-E^{\beta'}(\omega) {}^-E^{\gamma'}(\omega) \quad (S3.2)$$

Then we have:

$$\chi^{(2)}_{\alpha'\beta'\gamma'} = \sum_{\alpha,\beta,\gamma} {}^-R_\alpha^{\alpha'} \chi^{(2)}_{\alpha\beta\gamma} \left[{}^+R_\beta^{\beta'}\right]^{-1} \left[{}^+R_\gamma^{\gamma'}\right]^{-1} \quad (S3.3)$$

Therefore, for the co-polarized SH conversion process, we have:

$$\chi^{(2)}_{R'L'L'} = {}^-R_R^{R'} \chi^{(2)}_{RLL} \left[{}^+R_L^{L'}\right]^{-1} \left[{}^+R_L^{L'}\right]^{-1} = \chi^{(2)}_{RLL} \exp(j\theta_{eff}) \quad (S3.4a)$$

$$\chi^{(2)}_{R'L'R'} = {}^-R_R^{R'} \chi^{(2)}_{RLR} \left[{}^+R_L^{L'}\right]^{-1} \left[{}^+R_R^{R'}\right]^{-1} = \chi^{(2)}_{RLR} \exp(-j\theta_{eff}) \quad (S3.4b)$$

$$\chi^{(2)}_{R'R'R'} = {}^-R_R^{R'} \chi^{(2)}_{RRR} \left[{}^+R_R^{R'}\right]^{-1} \left[{}^+R_R^{R'}\right]^{-1} = \chi^{(2)}_{RRR} \exp(-j3\theta_{eff}) \quad (S3.4c)$$

For the cross-polarized SH conversion process, we have:

$$\chi^{(2)}_{L'L'L'} = {}^-R_L^{L'} \chi^{(2)}_{LLL} \left[{}^+R_L^{L'}\right]^{-1} \left[{}^+R_L^{L'}\right]^{-1} = \chi^{(2)}_{LLL} \exp(j3\theta_{eff}) \quad (S3.5a)$$

$$\chi^{(2)}_{L'L'R'} = {}^-R_L^{L'} \chi^{(2)}_{LLR} \left[{}^+R_L^{L'}\right]^{-1} \left[{}^+R_R^{R'}\right]^{-1} = \chi^{(2)}_{LLR} \exp(j\theta_{eff}) \quad (S3.5b)$$

$$\chi^{(2)}_{L'R'R'} = {}^-R_L^{L'} \chi^{(2)}_{LRR} \left[{}^+R_R^{R'}\right]^{-1} \left[{}^+R_R^{R'}\right]^{-1} = \chi^{(2)}_{LRR} \exp(-j\theta_{eff}) \quad (S3.5c)$$

Because $\theta_{eff} = 3\theta$, then for the co-polarized SH conversion process, we have:

$$\chi^{(2)}_{R'L'L'} = \chi^{(2)}_{RLL} \exp(j3\theta), \chi^{(2)}_{R'L'R'} = \chi^{(2)}_{RLR} \exp(-j3\theta), \chi^{(2)}_{R'R'R'} = \chi^{(2)}_{RRR} \exp(-j9\theta) \quad (S3.6)$$

For the cross-polarized SH conversion process, we have:

$$\chi^{(2)}_{L'L'L'} = \chi^{(2)}_{LLL} \exp(j9\theta), \chi^{(2)}_{L'L'R'} = \chi^{(2)}_{LLR} \exp(j3\theta), \chi^{(2)}_{L'R'R'} = \chi^{(2)}_{LRR} \exp(-j3\theta) \quad (S3.7)$$

Then the diffraction order observed in the measurement could accordingly correspond to the above nonlinear conversion processes.

Figure S6 shows the measured diffraction pattern of co-polarized and cross-polarized SH signals generated by the C3 silicon meta-atom placed in a square lattice, where $0^{th}$-order diffraction is removed to increase the visibility of the measurement. Here, the gradient metasurface made of C3 silicon meta-atom contains the periodic supercell made of 24 units (same sample as that shown in Figure 5(a)). It is apparent that the cross-polarized SH energy is mainly diffracted to $1^{st}$ order while co-polarized SH energy is diffracted to $\pm 1^{st}$ order. Considering the fundamental wavelength of 1200 nm, the measured diffracted angle of cross-polarized SH emissions is $-3.6°\pm 0.1°$, which is close to the value calculated by the cross-polarized SH conversion process $\chi^{(2)}_{L'L'R'} = \chi^{(2)}_{LLR} \exp(j3\theta)$, i.e., $-3.68°$. Similarly, the measured diffracted angle of the $\pm 1^{st}$ order of the co-polarized SH emissions is $-3.6°\pm 0.1°$ and $3.5°\pm 0.1°$, respectively, which are close to the theoretical value calculated by the SH conversion process $\chi^{(2)}_{R'L'L'} = \chi^{(2)}_{RLL} \exp(j3\theta)$ and $\chi^{(2)}_{R'L'R'} = \chi^{(2)}_{RLR} \exp(-j3\theta)$.

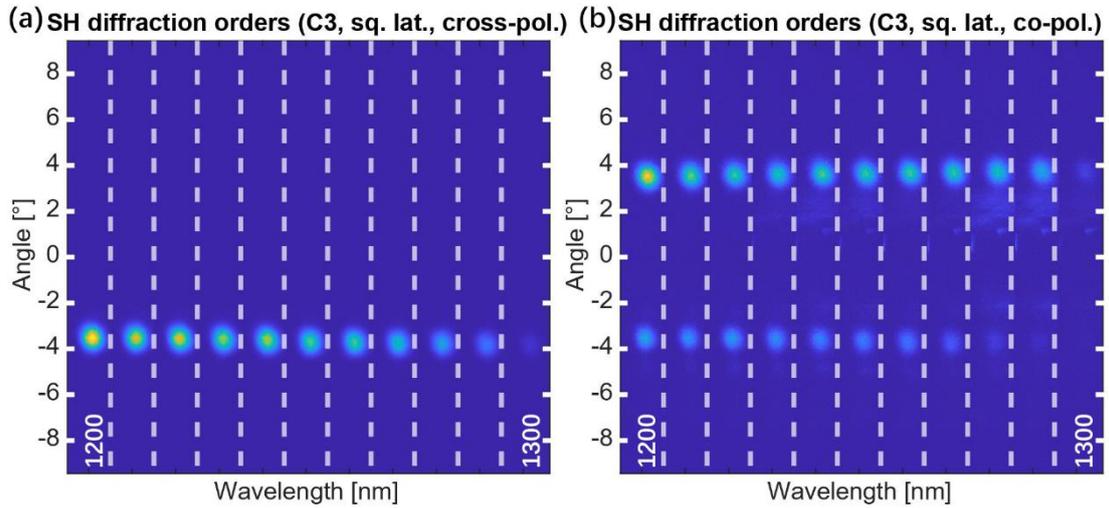

**Figure S6.** Experimental measurement of SHG of C3 silicon meta-atom array in square lattice. (a) Co-polarized and (b) Cross-polarized SH diffraction pattern when varying the fundamental wavelength from 1200 nm to 1300 nm by step of 10 nm.

## 5. Sample fabrication details

In this section, we show the detailed process of sample fabrication. See Figure S7. 3 nm of chromium (Cr), 100 nm of silver (Ag) and another 3 nm of chromium are deposited on a glass substrate using electron beam evaporation. The thin chromium layer is used to increase the adhesion of silver to the glass substrate as well as the adhesion of silicon to silver. After that, 105 nm of amorphous silicon (-Si) is deposited via plasma-enhanced chemical vapor deposition (PECVD) using a mixture of silane ($SiH_4$) and argon (Ar) at a temperature of 300°C. For lithography, 100 nm of the negative-tone resist ma-N 2401 is spin-coated and 40 nm of Elektra 92 is also added to increase the conductivity of the resist film. After patterning, the conductive layer is removed using water, and the patterned resist is developed for ~17 s in ma-D 525. The developed resist is hard-baked at 100° for 10 minutes to increase the etch stability. For etching a pseudo Bosch process is used with sulfur hexafluoride ($SF_6$) as an etch gas and perfluorocyclobutane ($C_4F_8$) as a passivation gas. After etching an estimated 20 nm of the etch mask remains and since removal via oxygen plasma oxidizes the silver, the thin resist residuals are left on top of the meta-atoms.

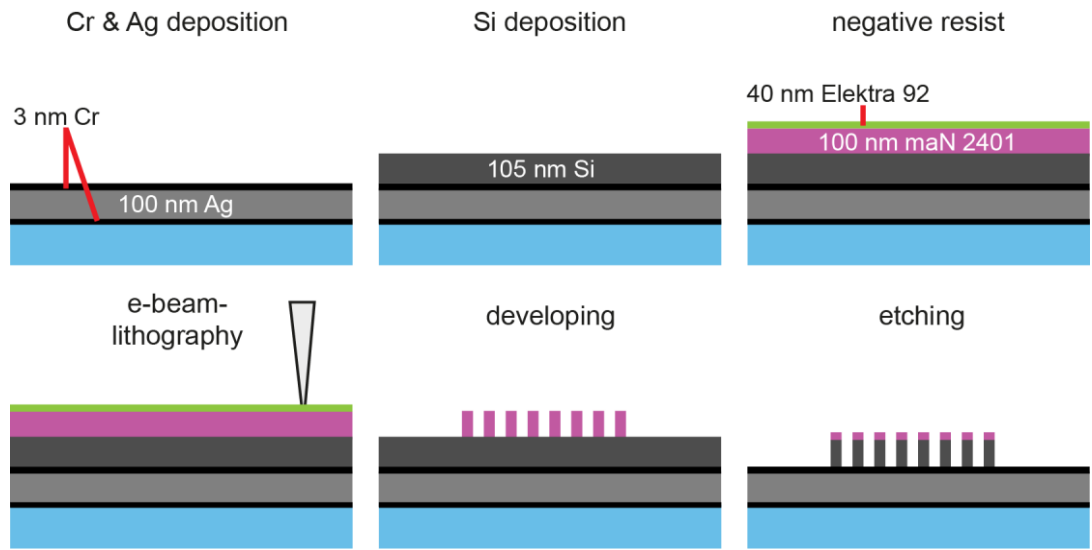

**Figure S7.** Flow chart of the sample fabrication.

## 6. Discussion on the experiment setup optimization

The dichroic mirror and off-axis lenses as well as other optical components in the experiment would change the ellipticity of the beams due to the different reflection amplitudes and phases for s- and p-polarized components of light, therefore, we should carefully design the experiment setup to minimize this effect.

Figure S8 (a) shows a setup by using a dichroic mirror in the measurement. Here,

a pulsed laser of 200 fs pulse width is generated with an optical parameter oscillator and functions as FW light source. Next, the FW wave passes through a quarter waveplate and is focused onto the sample metasurface (MS) via the reflection of a dichroic mirror (DM). TH emissions are collected with three focusing lenses, and then pass through the polarization analyzer set (a quarter waveplate and a linear polarizer) and a short pass filter. First, we test the polarization state of the fundamental wave. We noticed a wavelength-dependent distortion of the polarization state after reflection off the dichroic mirror leading to strong ellipticity for some wavelengths. We measured the polarization state of the fundamental beam after the dichroic mirror by using a linear polarizer and a power meter. Figure S8(b) shows such a measurement for a wavelength of 1200 nm, where we expect a constant power as a function over rotation angle, however, the result shows a nearly linear polarized state. This polarization distortion explains the "beating" effect in the measured TH diffraction patterns, where the diffraction orders shift from one half of the k-space to the other when changing the excitation fundamental wavelength, see Figure S8 (d,e).

Besides the dichronic mirror, we noticed that other beam splitters also introduced a similar effect and to avoid using a beam splitter, we eventually used a silver mirror with a small angle of incidence to minimize the polarization distortion. The lens used to focus on the metasurfaces is used slightly off-axis to generate a beam-offset between the incident fundamental wave and the generated third harmonic. Because of this off-axis use, the sample is illuminated under an angle of ~2°. This incident angle allows the generated TH to pass the silver mirror. The corresponding experiment setup can be found in Figure 4(a) of the main text. We repeated the measurements from Figure S8(a) for the new setup at different wavelengths as well as for the THG to verify the circular polarization state. One reference measurement for a wavelength of 1240 nm is done behind the quarter-waveplate and before the FW reflects off the mirror, see Figure S8(c). The ratio between the long and short semi-axis of the FW's elliptical polarization state is ~1.26 and nearly constant over wavelength. The TH shows a similar behavior albeit with a slightly larger power variation over the set linear polarization. The improvement can be seen from the measurement shown in Figure 5(c,d) of the main text.

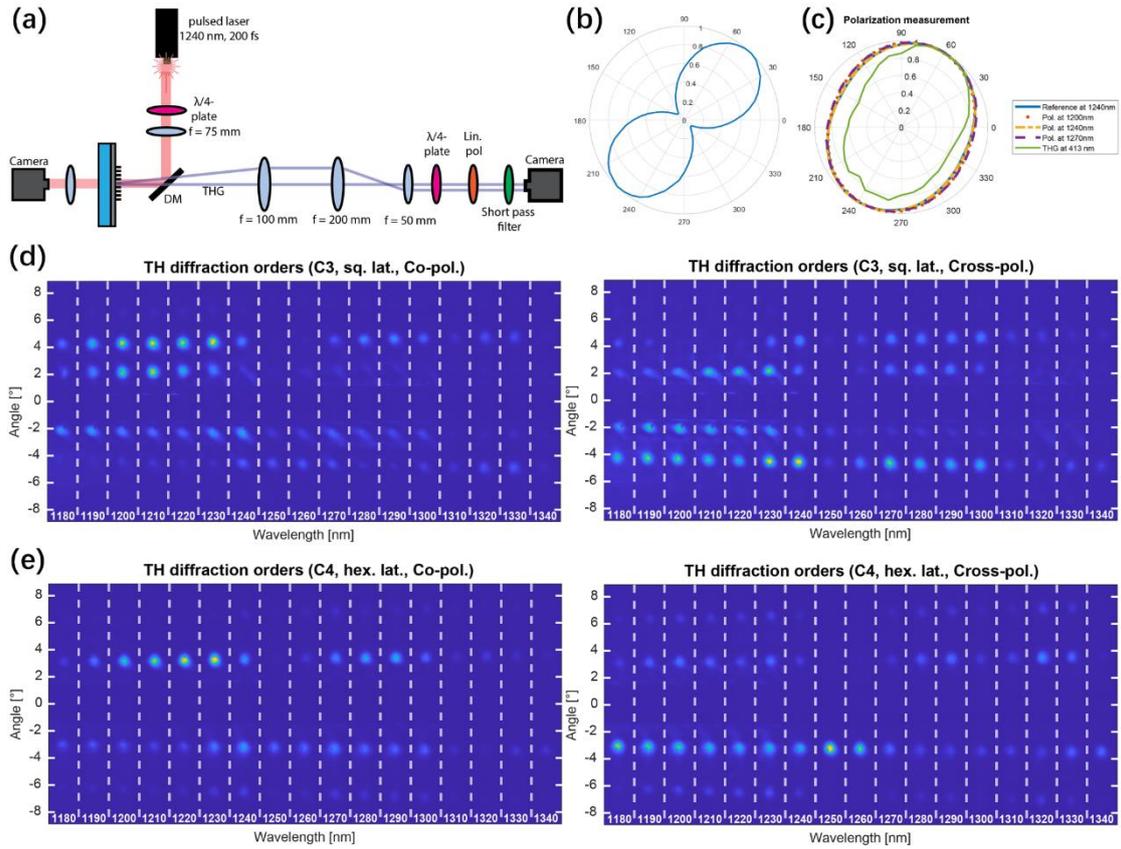

**Figure S8.** (a) Experiment setup by using dichroic mirror. (b) Polarization state of the fundamental beam at 1200 nm. (c) Polarization state of the fundamental and the third harmonic wavelength with the experiment setup in the main text Figure 4(a). 'Beating' effect of measured TH diffraction pattern when utilizing the dichronic mirror, the co- and cross-polarized TH emissions from the sample made of (d) C3 meta-atoms placed in square lattice and (e) C4 meta-atoms placed in hexagonal lattice when varying the fundamental wavelength from 1180 nm to 1340 nm in steps of 10 nm.